\documentclass[journal]{IEEEtran}

\IEEEoverridecommandlockouts 

\usepackage{blindtext, graphicx}
\usepackage{booktabs}
\usepackage[cmex10]{amsmath}
\usepackage{mathtools}
\usepackage{nicefrac}

\usepackage[small,font=scriptsize]{caption}

\usepackage{url}
\usepackage{graphics}
\usepackage{amssymb}
\usepackage{siunitx}
\usepackage{hyperref}
\usepackage{geometry}
\usepackage{todonotes}
\usepackage{rotating}
\usepackage{colortbl}
\usepackage{setspace}

\let\labelindent\relax
\usepackage{enumitem}

\usepackage[ruled,vlined]{algorithm2e}

\usepackage{multirow}
\usepackage{xcolor}
\usepackage[multiple]{footmisc}

\newtheorem{lemma}{Lemma}
\newtheorem{theorem}{Theorem}
\newtheorem{proof}{Prof} 
\newtheorem{obs}{Observation}

\newcommand{\theModel}{CF\nobreakdash-EVRP\xspace}
\newcommand{\theAlgorithm}{ComSat\xspace}
\newcommand{\MonoMod}{MonoMod\xspace}
\newcommand{\theRelaxedAlgorithm}{C\nobreakdash-ComSat\xspace}

\newcommand{\OR}{\ensuremath{\mathit{OR}}\xspace}
\newcommand{\C}{\ensuremath{\mathit{C}}\xspace}
\newcommand{\D}{\ensuremath{\mathit{D}}\xspace}
\newcommand{\serviceTime}{\ensuremath{\mathit{S}}\xspace}
\newcommand{\speed}{\ensuremath{\mathit{v}}\xspace}

\newcommand{\Origin}{\textrm{$ \mathcal{O}$}}
\newcommand{\OriginStartSet}{\textrm{$ \mathcal{S}$}}
\newcommand{\OriginFinishSet}{\textrm{$ \mathcal{F}$}}
\newcommand{\jobSet}{\ensuremath{\mathcal{J}}}

\newcommand{\taskSet}{\textrm{$\mathcal{K}$}\xspace}
\newcommand{\precTask}{\textrm{$\mathcal{P}$}}
\newcommand{\nodeSet}{\textrm{$\mathcal{N}$}}
\newcommand{\nodeset}[1]{\ensuremath{\mathcal{N}_#1}\xspace}
\newcommand{\vehicleSet}{\textrm{$\mathcal{V}$}}
\newcommand{\edgeSet}{\textrm{$\mathcal{E}$}}

\newcommand{\Perm}{\textrm{\emph{Perm}}}
\newcommand{\Start}{\textrm{\emph{s}}}

\newcommand{\CurrentRoutes}{\ensuremath{\mathit{CR}}\xspace}
\newcommand{\PreviousRoutes}{\ensuremath{\mathit{PR}}\xspace}

\newcommand{\CurrentAssignment}{\ensuremath{\mathit{CA}}\xspace}
\newcommand{\PreviousAssignment}{\ensuremath{\mathit{PA}}\xspace}

\newcommand{\ShortestPaths}{\ensuremath{\mathit{SP}}\xspace}
\newcommand{\CurrentPaths}{\ensuremath{\mathit{CP}}\xspace}
\newcommand{\PreviousPaths}{\ensuremath{\mathit{PP}}\xspace}
\newcommand{\newPaths}{\ensuremath{\mathit{NP}}\xspace}

\newcommand{\pathStart}{\textrm{$\xi$}\xspace}
\newcommand{\pathEnd}{\textrm{$\pi$}\xspace}
\newcommand{\incomingEdge}{\textrm{$\mathcal{I}$}}
\newcommand{\outgoingEdge}{\textrm{$\mathcal{U}$}}

\newcommand{\CapacityVerifySolution}{\ensuremath{\mathit{CVS}}\xspace}

\newcommand{\RoutesVerifyingFeasibility}{\ensuremath{\mathit{RVF}}\xspace}

\newcommand{\late}{\textrm{\emph{late}}}
\newcommand{\length}{\textrm{\emph{length}}}
\newcommand{\allo}{\textrm{\emph{$\alpha$}}}
\newcommand{\Finish}{\textrm{\emph{f}}}
\newcommand{\End}{\textrm{\emph{e}}}

\newcommand{\node}{\textrm{\emph{x}}}
\newcommand{\edge}{\textrm{\emph{y}}}
\newcommand{\NodeList}{\ensuremath{\mathit{NL}}\xspace}
\newcommand{\EdgeList}{\ensuremath{\mathit{EL}}\xspace}

\newcommand{\useNode}{\textrm{\emph{w}}}
\newcommand{\useEdge}{\textrm{\emph{z}}}

\newcommand{\timeHorizon}{\textrm{\emph{T}}}

\newcommand{\iteration}{\textrm{\emph{h}}\xspace}

\title{A Compositional Algorithm for the Conflict-Free Electric Vehicle Routing Problem}

\author{Sabino Francesco Roselli$^{1}$ and Per-Lage G\"otvall$^{2}$ and Martin Fabian$^{1}$ and Knut \AA kesson$^{1}$     
\thanks{We gratefully acknowledge financial support from Chalmers AI Research Centre (CHAIR) and AB Volvo (Project ViMCoR), and the Wallenberg AI, Autonomous Systems and Software program (WASP) funded by the Knut and Alice Wallenberg Foundation.

$^{1}$Department of Electrical Engineering, Chalmers University of Technology,
        G\"oteborg, Sweden
        {\tt\small \{rsabino, fabian, knut\} @chalmers.se}.
$^{2}$Senior Research Engineer at Volvo Group Trucks Operations {\tt\small Per-Lage.Gotvall@volvo.com} 
}}

\begin{document}

\maketitle

\begin{spacing}{1.15}

    \begin{abstract}
    The Conflict-Free Electric Vehicle Routing Problem (\theModel) is an extension of the Vehicle Routing Problem (VRP), a combinatorial optimization problem of designing routes for vehicles to visit customers such that a cost function, typically the number of vehicles or the total travelled distance, is minimized. The problem finds many logistics applications, particularly for highly automated logistic systems for material handling.
    The \theModel involves constraints such as time windows on the delivery to the customers, limited operating range of the vehicles, and limited capacity on the number of vehicles that a road segment can accommodate at the same time. In this paper, the compositional algorithm \theAlgorithm for solving the \theModel is presented. The algorithm iterates through the sub-problems until a globally feasible solution is found. The proposed algorithm is implemented using an optimizing SMT-solver and is evaluated against an implementation of a previously presented monolithic model.
    The soundness and completeness of the algorithm are proven, and it is benchmarked on a set of generated problems and found to be able to solve problems of industrial size.
    
    \end{abstract}
    \def\abstractname{Note to Practitioners}
    \begin{abstract}
The need to define and solve the  \theModel relates to an industrial application where a fleet of autonomous robots navigate in a heterogeneous environment, shared with humans and other vehicles and obstacles. 
To allow for a low-level trajectory controller to handle dynamic obstacles, like humans and fork-lifts, the \theModel includes capacity constraints on the road segments. This increases the problem complexity, and thus requires to trade off optimality for feasability; this so to get solutions in reasonable time with respect to 
how long ahead the jobs to schedule are known.
The overall problem is to find feasible solutions that satisfy all constraints while avoiding travelling unnecessarily long routes, and at the same time meet the stipulated time-windows to deliver material just-in-time. 
The compositional algorithm (\theAlgorithm) presented in this work is based on the idea to break down the overall scheduling problem into sub-problems that are easier to solve, and then to build a schedule based on the solutions of the sub-problems. 
\theAlgorithm is designed to work well for industrial scenarios where there are good reasons to believe that feasible solutions do exist. 
This seems a reasonable assumption as in an industrial setting a sufficient number of mobile robots can typically be assumed to be available.
    \end{abstract}

\end{spacing}

\section{Introduction}

The use of mobile robots for \emph{just-in-time} deliveries is of considerable interest for modern manufacturing facilities~\cite{azadeh2017robotized}. The problem treated in this paper originates from an industrial need to use a fleet of Automated Guided Vehicles (AGVs) that run through the plant to deliver components to workstations just-in-time for them to be used. 
In this scenario, the scheduler needs to consider several types of constraints in addition to the time constraints. (i) AGVs have a limited operating range and need to recharge their battery when the state-of-charge becomes low. (ii) Jobs have specific requirements on the AGV to execute them where only some AGVs can handle some jobs. (iii) The plant layout may limit the AGV's freedom of movement; for instance, a passage may not be large enough to accommodate more than a fixed number of AGVs at a time.
Thus, we define a \emph{capacity} of the road segments, intersections, and workstations and include \emph{capacity constraints} in the problem formulation. A schedule is said to be  \emph{conflict-free} if it fulfills the capacity constraints at all times.

The constraints discussed above substantially increase the complexity of the problem, and even finding feasible solutions can take an unreasonable amount of time.
For example, a solution to a standard VRP can be computed within minutes for up to 100 customers and just as many vehicles~\cite{zhang2021review}. On the other hand, when additional features such as charging times, capacity constraints, and multiple assignments of routes to vehicles come into play, a problem involving 10 vehicles and 20 customers can be considered large (plus the size of the plant also becomes a parameter of the problem), and it may take  hours to find a feasible solution for it. 
Moreover, in industrial applications the most essential requirement is delivering the goods within time windows to avoid delays in the production. For these reasons, the goal of the algorithm presented in this work is to find feasible solutions, rather than necessarily optimal ones.

There exist both approximate and exact algorithms to solve the VRP. For relatively small-size problem instances, mixed-integer linear programming  (MILP)~\cite{brahimi2016multi} solvers can find feasible, or even optimal, solutions in a relatively short time. However, standard MILP-solving techniques do not scale well, so specific-purpose algorithms involving local search~\cite{braysy2005vehicle}, Benders decomposition~\cite{riazi2020energy}, or stochastic methods~\cite{baker2003genetic, gong2011optimizing} have been proposed. Recent work focusing on fleets of electric vehicles~\cite{rossi2019interaction}, as well as conflict-free routing~\cite{thanos2019dispatch} show applications of such approaches to real-world problems.

In \cite{jernheden2020comparison} a comparison of using MILP as an exact method, and a set-based particle-swarm optimization algorithm as an approximate method, is made for solving the VRP with time windows (VRPTW~\cite{desrochers1992new}). The comparison shows that neither method dominates the other in terms of running time and quality of the solutions. On the other hand, the advantage of using general-purpose MILP-solvers is that additional constraints can be easily added to handle extensions of the original problem. At the same time this may be non-trivial, if at all possible, for a tailor-made algorithm.

The specific scheduling and routing problem treated in this paper, called the \emph{Conflict-Free Electric Vehicle Routing Problem} (\theModel), does involve additional constraints such as limited operating range of the robots, and capacity constraints, thus a general-purpose solver is used. 
In~\cite{roselli2018smt, roselli2019smt} optimizing Satisfiability Modulo Theory (SMT~\cite{barret_2009, DeMoura_2011}) solvers outperformed MILP solvers on combinatorial scheduling problems such as Job Shop Problems (JSP) involving many logical constraints. The natural abilities of SMT solvers to natively handle combinatorial constraints make them well suited to handle \theModel.

A monolithic model of the \theModel is presented in~\cite{roselli2021smt} and solved using the SMT-solver Z3~\cite{bjorner2015nuz}.
Already relatively small systems, with only a few vehicles and jobs, result in hundreds of thousands of variables and constraints due to the discretization of time used to model capacity constraints. Therefore, in~\cite{roselli2021compo_algo} a compositional algorithm was introduced that breaks down the \theModel into sub-problems such that discretization is avoided. The compositional algorithm scales better, in terms of computational cost, to larger models, but does not necessarily guarantee optimal solutions with respect to total travelled distance.

This work introduces \emph{\theAlgorithm} (Compositional Satisfiability), an extension of the compositional algorithm introduced in~\cite{roselli2021compo_algo}. The contributions in this paper are:
(i) a generalization of the \theModel so that vehicles can be located at multiple depots, and service times at the customers are accounted for;
(ii) a presentation of the extended compositional algorithm (now called \theAlgorithm) of~\cite{roselli2021compo_algo}, which can handle the generalized \theModel. Also, the search for alternative paths is improved by formulating it as Boolean satisfiability; 
(iii) a proof of \theAlgorithm's soundness and completeness under given restrictions;
(iv) an evaluation of \theAlgorithm on a set of generated instances of the \theModel.

The paper is organized as follows. Section~\ref{sec: literature_review} introduces previous works on the topic and puts this work in context. Section~\ref{sec: problem_definiton} provides a formal description of the problem. In Section~\ref{sec:compo_algo}, \theAlgorithm is introduced. Proof of soundness and completeness is given in Section~\ref{sec: proofs}. In Section~\ref{sec:experiments}, the results of the analysis over a set of problem instances are presented. Finally, conclusions are drawn in Section~\ref{sec:conclusions}.

\section{Literature Review} \label{sec: literature_review}

The VRP~\cite{dantzig1959truck} is a classical problem, formulated by Dantzig and Ramser, that searches for optimal routes for a fleet of robots to visit a set of customers. A large number of studies have introduced variations on the original problem, as well as techniques to solve them. 
The vehicle routing problem with time-windows (VRPTW) is an extension of the VRP where customers have to be visited within given time windows~\cite{desrochers1992new}. A related problem is discussed  in~\cite{smolic2009time} where the routes to visit customers are dynamically designed based on the current state of the other vehicles. 
Another variation of the problem involves the possibility of Multiple Depots (MDVRP). In~\cite{lim2005multi}, the MDVRP is decomposed into assignment of vehicles to customers, and design of routes for vehicles to visit their assigned customers. In~\cite{lau2009application}, the MDVRP is solved by means of genetic algorithms.  
In~\cite{krishnamurthy1993developing} the problem of limited capacity of the road segments is tackled and conflict-free routes for AGVs are computed by means of column generation.
The work in \cite{correa2007scheduling} presents one of the first and most relevant works involving conflict-free routing in combination with scheduling of jobs for flexible manufacturing systems. Similarly to~\cite{lim2005multi}, the authors break down the problem into a scheduling problem, solved by constraint programming, where vehicles are assigned to jobs, and a routing problem, solved by MILP, where routes for the vehicles are designed. We took inspiration from this approach and further broke down the problem into more sub-problems to be able to handle the different constraints, in particular the limited operating range of the vehicles.

For electric vehicles, model formulations need to take into account the vehicle's limited operating range and non-negligible charging time as well. 
In \cite{bard1998branch} a branch and cut algorithm to solve a VRP with satellite facilities is presented, vehicles can stop to replenish their cargo and continue delivering goods until the end of their shift. Satellite facilities are also treated by~\cite{schneider2014electric}, that models these intermediate points as charging stations and solves the problem by means of a combination of neighborhood and tabu search.

Autonomous vehicles are increasingly used to deliver material across manufacturing plants. 
In~\cite{nishi2010petri}, AGVs are scheduled for jobs and routed through a plant by means of Petri net decomposition. In~\cite{mousavi2017multi}, a hybrid evolutionary algorithm is implemented to solve a multi-objective AGV scheduling problem in a flexible manufacturing system. In this work, the authors consider the vehicles' battery charge, but do not take into account road segments' capacity.
In~\cite{liu2019multi}, a multi-objective AGV scheduling problem is solved by means of adaptive-genetic algorithms. Unlike standard genetic algorithms, these adjust the hyperparameters, improving convergence accuracy and speed. The authors consider a plant with a grid-like road network, but road segment capacities are not considered.
In \cite{rahman2020integrated} an integrated approach to deal with line balancing and material handling by means of AGVs is presented. A stochastic algorithm is used to assign jobs to the workstations and AGVs are scheduled to deliver the components to execute the jobs. However, road segment capacity constraints are not considered.
In~\cite{singh2021matheuristic}, a matheuristic (a combination of metaheuristics and mathematical programming) to schedule a heterogeneous fleet of AGVs is presented, having different travel speed and cost, charging/discharging rate, and capability to serve different customers. But again, road capacity constraints are not treated.

As AGVs are used in manufacturing plants with limited capacity of the road segments, a growing attention has been paid to the problem of designing conflict-free routes. 
In~\cite{saidi2015ant}, an ant colony algorithm is applied to the problem of job shop scheduling and conflict free routing of AGVs. While road segment capacity constraints are considered the limited operating range of the vehicles is not considered. 
In~\cite{yuan2016research}, a collision-free path planning for multi AGV systems based on the $A^*$ algorithm is presented. In this work, the environment is modeled as a grid, and conflicts can originate from vehicles occupying the same spot on the grid at the same time; the vehicles' operating range and ability to recharge is not considered. 
In~\cite{thanos2019dispatch}, a heuristic approach to solve the conflict-free routing problem with storage allocation is presented; in this work limited operating range and battery charge are not considered.
In~\cite{murakami2020time}, a MILP formulation to design conflict-free routes for capacitated vehicles is presented. This is an exact method, but it can only solve relatively small problem instances.
In \cite{zhong2020multi} a hybrid evolutionary algorithm to deal with conflict-free AGV scheduling in automated container terminals is presented. In this work, only a limited portion of the map is prone to conflicts, with all road segments allowing to travel in both directions simultaneously; also, charging of the vehicles' batteries is not considered. 
In~\cite{rizkallah2020smt} the authors present a new model formulation for the VRPTW that restricts the problem to only difference logic (a fragment of linear arithmetic) constraints, in order to exploit the strength of Z3 in dealing with this particular fragment. 

From the literature review it emerges that over the last twenty years there has been a growing number of studies dedicated to AGV-based material handling systems. There is usually a large overlap of features tackled in each work, and approximate methods are likely to be used to solve large problem instances, due the problem being too complex to be solved by exact algorithms.
We apply a graph-based concept similar to~\cite{yuan2016research}, with nodes and edges of the graph representing the plant. And similarly to~\cite{rizkallah2020smt} we do exploit the strength of Z3 in dealing with difference logic by turning the \emph{Assignment Problem} and the \emph{Capacity Verification Problem} (see Section~\ref{sec:compo_algo}) into JSPs, that can be described using difference logic.
However, to the best of our knowledge, there is no work that tackles at the same time both the limited operating range, with the necessity to recharge the vehicles' batteries, and the limited capacity of the road segments, requiring to schedule conflict-free routes. Moreover, there is no work using SMT solvers to compute schedules for multi-AGV systems comparable to our own.

\section{Problem Definition and Notation} \label{sec: problem_definiton}

\begin{figure*}[ht]
    \centering
    \includegraphics[width=0.9\textwidth]{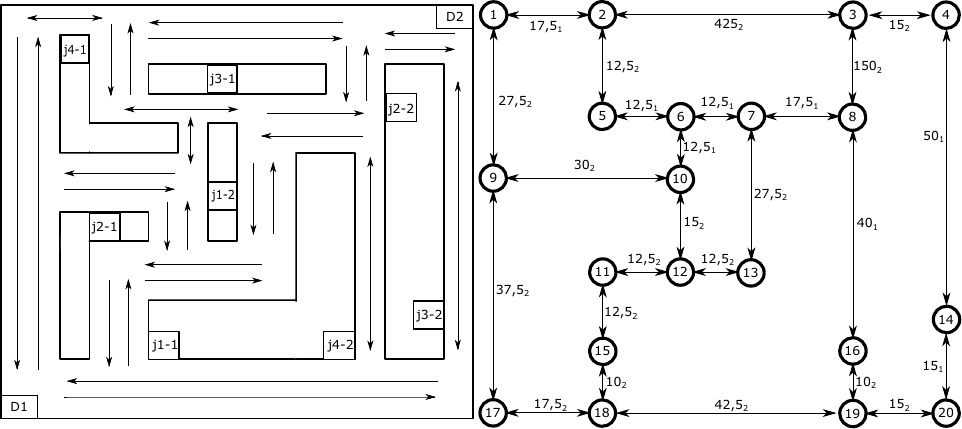}
    \caption{Problem instance of the \theModel picturing a hypothetical plant (left) where two depots ($D1$, $D2$) accommodate four vehicles ($v1$, $v2$, and $v3$, $v4$ respectively), available to execute four jobs ($j1$, $j2$, $j3$, $j4$), each composed by two tasks ($1$, $2$). The plant road segments are abstracted into a strongly connected, directed, weighed graph (right).}
    \label{fig:example}
\end{figure*}

In the \theModel the plant layout is represented by a finite, strongly connected, weighted, directed graph, where edges represent road segments and nodes represent either intersections between road segments or locations of customers.
A customer is defined by a unique (numerical) identifier, a location, and a time window, i.e., a lower and an upper bound that represent the earliest and latest arrival time allowed to serve the customer. The word \emph{customer} is typically used to denote a location where a vehicle is to pick up or drop off goods. The word \emph{task} will be used synonymously.
Edges have two attributes representing a road segment's length and its capacity in terms of number of vehicles that it can simultaneously accommodate.

The following definitions are provided:
\begin{itemize}

    \item Node: a location in the plant. 
    A node can only accommodate one vehicle at a time unless it is a \emph{hub} node.
    \begin{itemize}[label={}]
        \item \nodeSet: a finite set of nodes.
        \item $\nodeset{H} \subseteq N $: the set of hub nodes, nodes that can accommodate an arbitrary number of vehicles.
    \end{itemize}
    
    \item Edge: a road segment that connects two nodes.
    \begin{itemize}[label={}]
        \item $\edgeSet \subseteq \nodeSet \times \nodeSet$: the finite set of edges.
        \item $\bar{e}, \ \forall e \in \edgeSet$: the reverse edge of an edge $e$.
        \item $ d_{nn'} \in \mathbb{R}, \ \forall \langle n,n'\rangle \in \edgeSet$: the length of the edge connecting nodes $n$ and $n'$.
        \item $ g_{nn'} \in \{1,2\}, \ \forall \langle n,n' \rangle \in \edgeSet$: the capacity of the edge connecting nodes $n$ and $n'$
    \end{itemize}
    \item Time horizon: a fixed point of time in the future when all jobs are assumed to have ended.
    \begin{itemize}[label={}]
        \item $ \timeHorizon $: the time horizon. 
    \end{itemize}
    
    \item Job: a set of tasks that must be executed within the same route and without executing any task belonging to a different job in between. Typically, a job is the pickup of parts from the warehouse and the delivery to the due workstation; however, in general a job can have an arbitrary number of tasks. No task belonging to another job should be executed in the same route until all tasks belonging to the current job are completed. 
    
   \begin{itemize}[label={}]
        \item  \jobSet: the finite set of jobs.
   \end{itemize}
    
    \item Task: either a pickup or a delivery operation (there is no need to distinguish between them, as both are modelled in the same way). A task is always associated with a node where it is executed, and has a time window indicating the earliest and latest time at which it can be executed. Unless explicitly given, the time window spans the entire time horizon $[0,T]$. Also, each task is associated with a precedence list that states what tasks have to be executed before it. This list may include any other task in the problem definition. Finally, for each task a service time is defined. 
    \begin{itemize}[label={},leftmargin=*]
        \item $\taskSet$
        the finite set of all tasks.
        \item $\taskSet_j\subseteq\taskSet, \ \forall j \in \jobSet $ : the finite set of tasks of job $j$. (note that the tasks set is partitioned into subsets based on the jobs, i.e., $\taskSet_i \cap \taskSet_j = \emptyset \ \forall i \neq j, \ i, j \in \jobSet $)  
        \item $ L_{k} \in \nodeSet, \ \forall j \in \jobSet , \ k \in \taskSet_j $ : the location of task $k$. 
        \item $ \precTask_{k} \subset \taskSet_j, \ \forall j \in \jobSet, \ k \in \taskSet_j $: the set of tasks to execute before task $k$.
        \item $ l_{k} \in \mathbb{R}, \ \forall j \in \jobSet, \ k \in \taskSet_j $: the time window's lower bound for task $k$.
        \item $ u_{k} \in \mathbb{R}, \ \forall j \in \jobSet, \ k \in \taskSet_j $: the time window's upper bound for task $k$.
        \item $ \serviceTime_{k} \in \mathbb{R}, \ \forall j \in \jobSet, \ k \in \taskSet_j $: the service time of task $k$. 
    \end{itemize}
    
    \item Depots: nodes at which one or more vehicles start and must return to after completing the assigned jobs. A depot can accommodate an arbitrary number of vehicles at the same time, thus all depots are hubs.
    \begin{itemize}[label={}]
        \item $\emptyset \subset \Origin \subseteq \nodeset{H}$: the set of       depots.
        \item $ \OriginStartSet = \{\,\Start_o \,|\, o \in \Origin \,\} $: the set of dummy tasks representing the start from depot $o$.
        \item $ \OriginFinishSet = \{\,\Finish_o \,|\, o \in \Origin \,\} $: the set of dummy tasks representing the arrival at depot $o$.
    \end{itemize}
    \noindent The sets \OriginStartSet~and \OriginFinishSet~are disjoint with each other and with all task sets $\taskSet_j, \ \forall j \in \jobSet$.
    
    \item Vehicle: a transporter, e.g. a mobile robot, that is able to move between locations in the plant and perform pickup and delivery operations. 
    \begin{itemize}[label={}]
        \item  \vehicleSet: the finite set of all vehicles.
        \item  $ \vehicleSet_j \subseteq \vehicleSet, \ \forall j \in \jobSet $: set of vehicles eligible for job $j$.
        \item $\OR \in \mathbb{R^+}$: the vehicles' maximum operating range (constant).
        \item $\C \in \mathbb{R^+}$: the charging coefficient (constant).
        \item $\D \in \mathbb{R^+}$: the discharging coefficient (constant).
        \item $\rho \in \mathbb{R^+}$: a coefficient to scale remaining charge into remaining operating range (constant).
        \item $\speed \in \mathbb{R^+}$: vehicle speed (constant) while moving.
    \end{itemize}
\end{itemize}
\newpage
The requirements of the problem are:
\begin{itemize}
    \item All jobs have to be completed; for a job to be completed a vehicle has to be assigned to it and visit the locations of the job's tasks according to the tasks' sequence and within their respective time windows.
    \item Vehicles are not allowed to arrive at the task's location before the time window's lower bound and wait there (many other VRP formulations, allow such waiting).
    \item Vehicles are powered by batteries with limited capacity but with the ability to recharge at the depots. It is assumed that state of charge increases proportionally to the time spent at the depot and decreases proportionally to the travelled distance. Also, vehicles travel at constant speed \speed or they are stationary.
    \item Multiple depots; vehicles have to return to the depot they were dispatched from and can only recharge their batteries there (without queuing).
    \item A non-empty subset of vehicles is eligible for each job.
    \item All vehicles have the same operating range and start at full charge; whenever they return to the depot they charge to full before becoming available again; 
    \item Two additional jobs are added for each depot: \emph{start} and \emph{end}; they are needed in the \emph{Routing Problem} to make sure that routes begin and end at the depot. Both \emph{start} and \emph{end} have only one task located at the depot they represent, service time equal to zero, and the entire time horizon as time window. 
    \item Road segment capacities constrain the number of vehicles a road segment can simultaneously accommodate. 
    \item Only (non-cyclic) \emph{paths}, that is, finite sequences of edges that join sequences of distinct vertices, are considered.
\end{itemize}

\subsection{Example of the \texorpdfstring{\theModel}{theModel}}\label{sec:example}

Fig.~\ref{fig:example} shows an example of the \theModel, where four AGVs are available to execute four jobs, each composed by two tasks (the squares distributed over the plant). Each task is marked by a numeric code where the first digit refers to the job and the second digit indicates the task number. On the right is shown how the plant layout is abstracted into a strongly connected, directed, weighted graph (more on this below). The nodes represent the intersections of road segments in the plant; if a task's location is close enough to an intersection, then the task will be assigned that location, otherwise a new node is added to the graph (e.g., \emph{Node\,14} for \emph{task\,j3-2}). Also, nodes \emph{4} and \emph{17} represent both an intersection between road segments and the depots. The numbers on the edges represent the segments' length (regular font), and their capacity (subscript). 

The problem, using the above defined notation, is then:
\begin{itemize}[label={}]
    \item $ \nodeSet = \{ 1,\ldots,20 \}, \ \nodeset{H} = \{4,17 \},  \Origin = \{4,17 \}$
    \item $ \edgeSet = \{(1,2),(2,3),(3,4),(5,6),(6,7),(7,8),(9,10),$
    \item $ (1,12),(12,13),(17,18),(18,19),(19,20),(1,9), $ 
    \item $ (2,5),(3,8),(6,10),(7,13),(8,16),(9,17),(11,15), $ 
    \item $ (15,18),(16,19),(14,21)  \} $
    \item $\jobSet = \{j1,j2,j3,j4 \}$
    \item $\taskSet = \{i1,i2 \ | \ \forall i \in \jobSet \}$
    \item $L_{j11} = 15, \ L_{j12} = 10, \ L_{j21} = 11, \ L_{j22} = 8, $
    \item $L_{j31}=6, \ L_{j32} = 14, \ L_{j41} = 2, \ L_{j42} = 16 $
    \item $ \precTask_{j11} = \emptyset, \precTask_{j12} = {j11}, \precTask_{j21} = \emptyset, \precTask_{j22} = {j21}$
    \item $ \precTask_{j31} = \emptyset, \precTask_{j32} = {j31}, \precTask_{j41} = \emptyset, \precTask_{j42} = {j41}$
    \item $l_{j11} = 0, \ l_{j12} = 70, \ l_{j21} = 0, \ l_{j22} = 300, $
    \item $l_{j31}=0, \ l_{j32} = 180, \ l_{j41} = 0, \ l_{j42} = 200 $
    \item $u_{j11} = \timeHorizon, \ u_{j12} = 120, \ u_{j21} = \timeHorizon, \ u_{j22} = 330, $
    \item $u_{j31}=\timeHorizon, \ u_{j32} = 250, \ u_{j41} = \timeHorizon, \ u_{j42} = 300 $
    \item $\serviceTime_{j11} = 10, \ \serviceTime_{j12} = 30, \ \serviceTime_{j21} = 20, \ \serviceTime_{j22} = 30, $
    \item $\serviceTime_{j31} = 10, \ \serviceTime_{j32} = 30, \ \serviceTime_{j41} = 30, \ \serviceTime_{j42} = 20, $
    \item $\vehicleSet = \{ v1,v2,v3,v4 \}$
    \item $\vehicleSet_{j1} = \{ v1, v2 \}, \ \vehicleSet_{j2} = \{ v1 \}, $
    \item $ \vehicleSet_{j3} = \{ v2, v4 \}, \ \vehicleSet_{j4} = \{ v3, v4 \} $
    \item $ \OR = 270, \C = 3, \ \D = 1, \ \rho = 1, \ \speed = 1 $
    \item $\timeHorizon = 500$
\end{itemize}

First, \theAlgorithm will compute the distance for each pair of nodes 
where either a task or a depot is located. Using the distances, as well as the speed of the vehicles, and all the other parameters, a set of routes \CurrentRoutes is computed. In this specific problem, a possible set of routes is:

\noindent\textbf{$R1$}: $D1$-$j11$-$j12$-$D1$; Total length: 135; latest start: 42.5; Eligible vehicles: $v1$, $v2$

\noindent\textbf{$R2$}: $D1$-$j21$-$j22$-$D1$; Total length: 220; latest start: 200; Eligible vehicles: $v1$

\noindent\textbf{$R3$}: $D2$-$j31$-$j32$-$D2$; Total length: 257.5; latest start: 102.5; Eligible vehicles: $v2$, $v4$

\noindent\textbf{$R4$}: $D2$-$j41$-$j42$-$D2$; Total length: 195; latest start: 195; Eligible vehicles: $v3$, $v4$

Note that for each route the latest start time is computed based on the strictest time window.
Subsequently vehicles are assigned to the routes and an actual start time for the route is given. A possible assignment for the current routes would be $R1$: $v2$; $R2$: $v1$; $R3$: $v4$; $R4$: $v3$. The start time of all routes is 0.

Finally, the routes are \emph{capacity checked}, and this also produces a node-by-node schedule, i.e., the arrival time of a vehicle at each node that is included in the route: 

\noindent {\bf $v1$:} $v1$-17:0;
$v1$-18:18.5;
$v1$-15:40.5;
$v1$-11:53;
$v1$-12:85.5;
$v1$-13:98;
$v1$-7:153;
$v1$-8:300;
$v1$-7:347.5;
$v1$-6:360;
$v1$-10:372.5;
$v1$-9:402.5;
$v1$-17:440;

\noindent {\bf $v2$:} $v2$-17:0;
$v2$-18:19.5;
$v2$-15:29.5;
$v2$-11:52;
$v2$-12:64.5;
$v2$-10:79.5;
$v2$-9:139.5;
$v2$-17:177;

\noindent {\bf $v3$:} $v3$-4:0;
$v3$-3:16;
$v3$-2:58.5;
$v3$-1:76;
$v3$-2:103.5;
$v3$-5:116;
$v3$-6:128.5;
$v3$-7:141;
$v3$-8:170.5;
$v3$-16:210.5;
$v3$-8:270.5;
$v3$-3:285.5;
$v3$-4:300.5;

\noindent {\bf $v4$:} $v4$-4:0;
$v4$-3:15;
$v4$-8:30;
$v4$-7:142;
$v4$-8:169.5;
$v4$-3:184.5;
$v4$-4:199.5;
$v4$-14:250;
$v4$-4:330.

If the charging coefficient $C$ is increased from 3 to 9, both $R1$ and $R2$ could be assigned to vehicle $v1$, since the charging time would be short enough to allow the vehicle to execute the first route, go back to the depot, recharge and execute the second route without breaking any time windows. In this case there would still be four routes but three vehicles would be enough to execute them.
Moreover, if the operating range of the vehicles were increased to 290, all customers could be served with only three routes:

\noindent \textbf{$R1$}: $D1$-$j21$-$j22$-$D1$; Total length: 220; latest start: 200; Eligible vehicles: $v1$

\noindent \textbf{$R2$}: $D2$-$j11$-$j12$-$j31$-$j32$-$D2$; Total length: 280; latest start: 100; Eligible vehicles: $v2$

\noindent \textbf{$R3$}: $D2$-$j41$-$j42$-$D2$; Total length: 195; latest start: 195; Eligible vehicles: $v3$, $v4$

The solutions presented in this section have been computed using \theAlgorithm; a discussion on the algorithm's solving procedure is provided in the next section, after the algorithm itself has been described.

\subsection{State Space Analysis}

Although the state space size of \theModel is not directly proportional to the solving time, analyzing the state space growth provides an idea of the complexity of the problem itself. 
The parameters needed to analyze the state space are the number of tasks $|\taskSet|$, the number of nodes $|\nodeSet|$ in the graph representing  the plant, the number of vehicles $|\vehicleSet|$, the time horizon \timeHorizon, and the operating range of the vehicles \OR. 
To be able to compute the state space size, the domains of the real valued variables are, in this analysis, restricted to integers.  

Based on the model formulation from~\cite{schneider2014electric}, if capacity constraints are relaxed, the size of the state space is upper bounded by
\begin{equation}\label{eq:statespace_no_capacity}
    OR^{|\vehicleSet|\cdot|\taskSet|}
    \cdot
    \timeHorizon^{|\vehicleSet|\cdot|\taskSet|}
    \cdot
    2^{|\vehicleSet|\cdot|\taskSet|^2},
\end{equation}
which can be rewritten as 
\begin{equation}\label{eq:statespace_no_capacity_rewritten}
    2^{|\vehicleSet|\cdot |\taskSet| \cdot (|\taskSet| + \log_2{(\OR\cdot\timeHorizon)}))}. 
\end{equation}

Based on the monolithic model of \theModel from~\cite{roselli2021smt}, if capacity constraints are considered, the size of the state space is upper bounded by
\begin{equation}\label{eq:statespace_capacity}
2^{|\vehicleSet|\cdot|\taskSet|} 
\cdot
2^{|\vehicleSet|\cdot|\taskSet|\cdot \timeHorizon}
\cdot
\OR^{|\vehicleSet|\cdot\timeHorizon}
\cdot 
2^{2\cdot|\vehicleSet|\cdot|\nodeSet|\cdot \timeHorizon},
\end{equation}
which can be rewritten as 
\begin{equation}\label{eq:statespace_capacity_rewritten}
2^{|\vehicleSet| \cdot ( |\taskSet|+\timeHorizon (2\cdot |\nodeSet| + \log_2{(\OR)} + |\taskSet| )  )}. 
\end{equation}

Assuming that, for a fixed number of vehicles there exists a correlation between the number of tasks and the time horizon, we can define $\beta \in [0,1]$ such that $|\taskSet| = \beta \cdot \timeHorizon$. The ratio between \eqref{eq:statespace_capacity_rewritten} and \eqref{eq:statespace_no_capacity_rewritten} then has a dominant factor of $2^{2\cdot|\vehicleSet|\cdot|\nodeSet|\cdot\timeHorizon}$ that arises due to the capacity constraints.
Even for small problem instances, this number can be very large.
A compositional approach that iteratively solves smaller sub-problems potentially avoids such state space explosion.



\section{Problem Decomposition and Solving Procedure} \label{sec:compo_algo}

This section describes how the \theModel is decomposed into sub-problems. The first step is a computation of the shortest paths to connect each pair of tasks. In the \emph{Routing Problem} the paths are used to compute routes that start and end at the depots and serve all tasks within their time windows. 
Once the routes are computed, the \emph{Assignment Problem} matches them with the vehicles, which determines their execution times.
Finally, the \emph{Capacity Verification Problem} checks the routes and their execution times against the capacity constraints.
It may be required to explore different paths to connect two tasks of a route; this is done by solving the \emph{Paths Changing Problem} that, based on the paths used so far, will find new unexplored paths to connect the tasks. When new paths are computed, the algorithm verifies whether the routes still meet the time windows by solving the \emph{Routes Verification Problem}. 
Table~\ref{tab:algorithms} summarizes inputs and outputs of the algorithms developed to solve each of the sub-problems presented in the following sections.

\subsection{Computation of Shortest Paths}

In the first step, the shortest path between any two tasks is computed. This is done because the \emph{Routing Problem} is a VRPTW plus additional constraints on tasks precedence and routes length; in a VRP there exist exactly one path between any two tasks. On the other hand, in the \theModel there may be several ways to travel from one task to another. 
Moreover, computing the shortest paths (instead of any \emph{non-cyclic} path) is required for the algorithm to be sound and complete (see Section~\ref{sec: proofs}).
The shortest paths \ShortestPaths are computed using Dijkstra's algorithm~\cite{dijkstra1959note} and assigned to the set \CurrentPaths. Since each task's location is reachable from any other task's location, a path always exists; also, in case more solutions with the same cost exist, these will be explored in the \emph{Paths Changing Problem}, if needed.

\subsection{Routing Problem} \label{subsec:routing_problem}

\begin{table}[htb]\setlength\lightrulewidth{0.3pt}
    \normalsize
  \centering
  \caption{Algorithms to solve the sub-problems of Section~\ref{sec:compo_algo}. 
  Given within parentheses next to the name of the algorithm is the problem that it solves.
  }
    \begin{tabular}{p{0.45\textwidth}}
    \toprule
    {\bf Router} ({\it Routing Problem})   \\
    \midrule
    \textbf{Input:} \CurrentPaths, \PreviousRoutes \\
    \textbf{Output:} \CurrentRoutes \\
    Define optimization problem using \eqref{eq:routing_cost_function}-\eqref{eq:previous_routes} \\
    Optimize and extract \CurrentRoutes from the solution\\
    \toprule
    {\bf Assign} ({\it Assignment Problem})  \\
    \midrule
    \textbf{Input:} \CurrentRoutes, \PreviousAssignment \\
    \textbf{Output:} \CurrentAssignment \\
    Define feasibility problem using \eqref{eq:exactly_one_assignment}-\eqref{eq:previous_assignments} \\
    Solve and extract \CurrentAssignment from the solution \\
    \toprule
    {\bf CapacityVerifier} ({\it Capacity Verification Problem})   \\
    \midrule
    \textbf{Input:} \CurrentAssignment \\
    \textbf{Output:} \CapacityVerifySolution \\
    Define feasibility problem using \eqref{eq:route_start}-\eqref{eq:edges_inverse} \\
    Solve and extract \CapacityVerifySolution from the solution \\
    \toprule
    {\bf PathsChanger} ({\it Paths Changing Problem})  \\
    \midrule
    \textbf{Input:} \PreviousPaths \\
    \textbf{Output:} \newPaths \\
    Define optimization problem using \eqref{eq:path_changing_cost_function}-\eqref{eq:previous_paths} \\
    Optimize and extract \newPaths from the solution \\
    \toprule
    {\bf RoutesVerifier} ({\it Routes Verification Problem})  \\
    \midrule
    \textbf{Input:} \CurrentRoutes, \newPaths \\
    \textbf{Output:} \emph{True} or \emph{False} \\
    Define feasibility problem using \eqref{eq:checker_domain}-\eqref{eq:checker_autonomy} \\
    Solve problem and return \emph{True} if feasible, else \emph{False} \\
    \bottomrule
    \end{tabular}%
  \label{tab:algorithms}%
\end{table}%

Solving the \emph{Routing Problem} means to find a set of routes, i.e., a sequence of tasks' locations that begins and ends at the same depot, such that every task is served within its time windows and the length of each route does not exceed the operating range. This way, once vehicles are assigned to routes in the \emph{Assignment Problem}, they can execute them without having to recharge.
Additionally, the routes have to meet the constraints on the tasks' precedence, as introduced above and described below. 
At this stage capacity constraints are not considered, nor is the actual plant layout. Also, upper bounds on the number of available vehicles are neglected, i.e., there can be more routes than vehicles, since one vehicle can be assigned to more than one route. 

With some abuse of notation we can define the distance between two arbitrary tasks' locations as $d_{k_1k_2}$ instead of $d_{L_{k_1}L_{k_2}}, \ \forall k_1 \in \taskSet_{j_1}, \ k_2 \in \taskSet_{j_2}, \ j_1,j_2 \in \jobSet$. Also, let $M_j$ be the set of mutually exclusive jobs for job $j$ (i.e. vehicles eligible for job $j$ are not eligible for any of the jobs in $M_j$ due to requirements on the vehicle type);
let $\Perm_j$ be the set of permutations of tasks belonging to job $j$, where each element \emph{ord} in $\Perm_j$ is an ordered list of tasks and let $k_{\textrm{next}}$ be the task coming after task $k$ in \emph{ord}.

The set of decision variables used to build the model for the \emph{Routing Problem} are:
\begin{itemize}[label={}]
    \item $\theta_{k_1k_2}$: Boolean variable that is true if a vehicle travels from the location of task $k_1$ to the location of task $k_2$, false otherwise.
    \item $\gamma_{k}$: non-negative real variable that models the arrival time of a vehicle at the location of task $k$.
    \item $\epsilon_{k}$: non-negative real variable that models the remaining charge of a vehicle when it arrives at the location of task $k$.
\end{itemize}

If the solution of the \emph{Routing Problem} turns out to be inconsistent with the vehicles' assignments or the capacity constraints, a new solution must be computed in order to find alternative routes for the same combination of paths. 
Therefore it is necessary to keep track of the combinations of routes that have already been generated so these can be ruled out when solving the \emph{Routing Problem} again. 
Let the optimal solution to the \emph{Routing Problem} found at iteration \iteration be $ \CurrentRoutes = \bigcup_{k_1,k_2 \in \taskSet}{ \{ \theta^*_{k_1k_2} \} } $, where $\theta^*_{k_1,k_2}$, $\forall k_1,k_2 \in \taskSet$, is the value of $\theta_{k_1,k_2}$ in the current solution; also, let \PreviousRoutes be the set containing the optimal solutions found until the $(\iteration-1)$-th iteration.

The following logical operators are used to express cardinality constraints \cite{sinz2005towards} in the sub-problems:
\begin{itemize}
    \item[] $\textrm{EN}(a,n):$ exactly $n$ variables in the set $a$ are true;
    \item[] $\textrm{If}(c,o_1,o_2):$ if $c$ is \emph{true} returns $o_1$, else returns $o_2$.
\end{itemize}
To shorten the notation we will write $\textrm{EN}_{m\in M}(m,n)$  to denote $ \textrm{EN}({\bigcup\limits_{m\in M}\{m\},n)}$.
The model formulation for the \emph{Routing Problem} is as follows:

\begin{flalign}
    &\min_{k \in \taskSet_{j}, \ j \in \jobSet, \ \Start \in \OriginStartSet} \sum{\textrm{If}(\theta_{\Start k}, 1, 0)} \label{eq:routing_cost_function} \\
    &\epsilon_{k} \cdot \rho \leq \OR, \ \ \quad \qquad \qquad \qquad \qquad   \forall k \in \taskSet_j, \ j\in \jobSet  \label{eq:domain}\\
    &\neg{\theta_{kk}}, \quad \qquad \qquad  \qquad \qquad \qquad \quad \ \forall k \in \taskSet_j, \ j  \in \jobSet \label{eq:not_travel_same_spot} \\
    &\neg{\theta_{k \Start}}, \quad \qquad \qquad \qquad \qquad \ \forall k \in \taskSet_{j}, \ s \in \OriginStartSet, \ j \in \jobSet \label{eq:no_travel_to_start} \\
    &\neg{\theta_{\Finish k}}, \quad \qquad \qquad \qquad \qquad \forall f \in \OriginFinishSet, \ k \in \taskSet_{j}, \ j \in \jobSet \label{eq:no_travel_from_end} \\
    &\theta_{k_1k_2} \implies \gamma_{k_2} \geq \gamma_{k_1} + \serviceTime_{k_1} + {d_{k_1k_2}}/{v},  \nonumber \\
    &\qquad \qquad \qquad \qquad \forall k_1 \in \taskSet_{j_1}, \ k_2 \in \taskSet_{j_2}, \ j_1,j_2 \in \jobSet \label{eq:infer_arrival_time} \\
    &\theta_{k_1k_2} \implies \epsilon_{k_2} \leq \epsilon_{k_1} - D \cdot d_{k_1k_2}/{v},    \nonumber \\
    &\qquad \qquad \qquad \qquad \forall k_1 \in \taskSet_{j_1}, \ k_2 \in \taskSet_{j_2}, \ j_1,j_2 \in \jobSet \label{eq:autonomy} \\
    &\textrm{EN}_{\substack{k_2 \in \taskSet_{j_2}, \\ j_2 \in \jobSet}}{(\theta_{k_1k_2},1)}, \ \quad \forall k_1 \in \taskSet_{j_1}, \ j_1 \in \jobSet, \ j_1 \neq j_2 \label{eq:one_arrival}\\
    &\textrm{EN}_{\substack{k_1 \in \taskSet_{j_1} \\ j_1 \in \jobSet}}{(\theta_{kk_1},n)} \implies \textrm{EN}_{\substack{k_2 \in \taskSet_{j_2} \\ j_2 \in \jobSet}}{(\theta_{k_2k},n)}, \nonumber \\
    &\qquad \qquad \qquad \qquad \ \forall j \in \jobSet, \ k \in \taskSet_{j}, \ n = 1,\dots,|\jobSet| \label{eq:flow} \\
    &\textrm{EN}_{\substack{k_1 \in \taskSet_{j_2} \\ j_2 \in \jobSet}}{(\theta_{k_1k},n)} = \textrm{EN}_{\substack{k_2 \in \taskSet_{j_2} \\ j_2 \in \jobSet}}{(\theta_{kk_2},n)}, \nonumber \\
    &\qquad \qquad \qquad \qquad \quad \ \forall k \in \OriginStartSet \cup \OriginFinishSet, \ n = 1,\dots,|\jobSet| \label{eq:route_continuity} \\
    &\gamma_{k} \geq l_{k} \wedge \gamma_{k} \leq u_{k}, \quad \qquad \qquad \quad \ \forall k \in \taskSet_{j}, \ j \in \jobSet \label{eq:routing_time_window} \\
    &\neg{\theta_{k_1k_2}},  \quad \forall k_1 \in \taskSet_{j_1}, \ k_2 \in \taskSet_{j_2}, \ j_1 \in \jobSet, \ j_2 \in M_{j_1} \label{eq:mutual_exclusive} \\
    &\bigvee_{ord \in \Perm_j}{\left( \bigwedge_{k \in ord}{\theta_{kk_{\textrm{next}}}}\right)}, \ \ \quad \qquad \qquad \qquad  \forall j \in \jobSet \label{eq:order_within_job} \\
    &\bigwedge_{k' \in \precTask_{k}}{\gamma_{k} \geq \gamma_{k'}}, \ \qquad \qquad \ \qquad \quad \qquad \qquad \forall k \in \taskSet \label{eq:routing_precedence} \\
    &\bigvee_{\theta_{k_1k_2} \in \lambda}{\neg{\theta_{k_1k_2}}}, \qquad \qquad \qquad \qquad \qquad \quad \forall \lambda \in \mathit{PR}  \label{eq:previous_routes}
\end{flalign}

 The cost function to minimize \eqref{eq:routing_cost_function} is the total number of routes. This is done by minimizing the number of direct travels from the tasks representing the depots; \eqref{eq:domain} restricts the remaining charge to be lower than or equal to the maximum operating range; \eqref{eq:not_travel_same_spot} forbids to travel from and to the same location; \eqref{eq:no_travel_to_start} and \eqref{eq:no_travel_from_end} express that a vehicle can never travel to the start, nor travel from the end: \emph{start} and \emph{end} referring to the same depot are physically located at the same node, but they play different roles in the \emph{Routing Problem}, hence two different tasks; \eqref{eq:infer_arrival_time} constrains the difference in arrival time based on the distance for a direct travel between two points; \eqref{eq:autonomy} models the decrease of charge based on the distance travelled. For \eqref{eq:infer_arrival_time} and \eqref{eq:autonomy} distances are computed using the current paths \CurrentPaths; \eqref{eq:one_arrival} expresses that each task's location must be visited exactly once; \eqref{eq:flow} guarantees the flow conservation between start and end; \eqref{eq:route_continuity} ensures that all vehicles leaving the depots return after visiting the tasks' locations; \eqref{eq:routing_time_window} enforces the time on the routes; \eqref{eq:mutual_exclusive} expresses that there cannot be direct travel among mutual exclusive jobs. This constraint does not always guarantee that mutually exclusive jobs will never be executed in the same route. Some corner cases are not covered, but the inconsistency will be spotted in the \emph{Assignment Problem} so there is no need to further complicate the constraint (there would be need to enumerate a large number of task sequences and rule out the inconsistent ones by adding one constraint for each of them), since it would slow down the whole sub-problem solution; \eqref{eq:order_within_job} expresses that if a number of tasks belong to one job, they have to take place in sequence; \eqref{eq:routing_precedence} guarantees that precedence constraints among tasks are enforced. 
 Constraint~\eqref{eq:previous_routes} allows to rule out the previously computed sets of routes as a solution. This is necessary as this optimization sub-problem may be called multiple times during the execution of \theAlgorithm.

Based on the model described above, the algorithm \emph{Router} is defined, that takes the set of current paths \CurrentPaths and the set \PreviousRoutes, and returns a set of routes \CurrentRoutes that have not been selected yet; if the problem is infeasible, \CurrentRoutes is empty.

\subsection{Assignment Problem}

The routes \emph{CR} are generated in the \emph{Routing Problem} based only on the time windows and on the vehicles' operating range. The \emph{Assignment Problem} now allocates vehicles to the routes based on the actual availability of each type of vehicle.
Moreover, even though constraint~\eqref{eq:mutual_exclusive} partially prevents it, \emph{CR} may contain routes that involve mutually exclusive jobs and, while it would be possible to avoid this by adding additional constraints, it would be 
inconvenient to do so in the \emph{Routing Problem}, since there is no information about the vehicles assigned to the routes. On the other hand, once a set of routes is given, it is verified in the \emph{Assignment Problem} whether a vehicle is actually eligible for a route.

Therefore for each route $ r \in \CurrentRoutes $, we can define a list of jobs $\jobSet^{r} \subseteq \jobSet$ that are executed by the vehicle assigned to $r$, and the list of eligible vehicles for $r$, $El_{r} = \bigcap_{ j \in \jobSet^{r}}{\vehicleSet_j}$. Also, based on the time windows and service times 
of the jobs forming the routes, it is possible to work out the latest start of a route $\late_{r}$.
Since a route can include more than one job, the strictest time window will define the latest start for the route. Finally, for each route we can define the cumulative service time  $\serviceTime_r = \bigcup_{\substack{k \in j \\ j \in \jobSet^r}}{\serviceTime_k}$.

The \emph{Assignment Problem} is formulated as a JSP where routes are jobs (whose durations depend on their lengths $\length_r$) and vehicles are resources, with some additional requirements on the jobs' starting time. 
The set of decision variables used to build the model are:
\begin{itemize}[label={}]
    \item $\allo_{ir}$: Boolean variable that is true if vehicle $i$ is assigned to route $r$, false otherwise;
    \item $\Start_r$: non-negative real variable that models the start time of route $r$;
    \item $\End_r$: non-negative real variable that models the end time of route $r$.
\end{itemize}

It may be necessary to have different assignments for the same set of routes \CurrentRoutes, since two vehicles located at the same depot may have different states of charge and, therefore, lead to different outcomes when solving the \emph{Capacity Verification Problem}. 
Thus, let the optimal solution to the \emph{Assignment Problem} found at iteration \iteration be $ \CurrentAssignment = \bigcup_{ \substack{ i \in \vehicleSet \\ r \in \CurrentRoutes } } { \{ \allo^*_{ir} \} }$, where $\allo^*_{ir}$, $\forall i \in \vehicleSet, \, r \in \CurrentRoutes$, is the value of $\allo_{ir}$ in the current solution; also, let \PreviousAssignment contain the optimal solutions found until the $(\iteration-1)$-th iteration.

The model formulation for the \emph{Assignment Problem} is:
\begin{align} 
    &\textrm{EN}_{i\in V}{(\allo_{ir},1)}, \qquad \qquad \qquad \qquad \qquad \  \forall r \in \CurrentRoutes \label{eq:exactly_one_assignment} \\
    &\End_r = \Start_r + {\length_r}/{\speed} + \serviceTime_r, \qquad \qquad \qquad    \forall r \in \CurrentRoutes \label{eq:end_based_on_length} \\ 
    &\Start_r \leq \late_r, \qquad \qquad \qquad \qquad \qquad \qquad \ \forall r \in \CurrentRoutes \label{eq:latest_start} \\
    &\bigvee_{i\in El_r}{\allo_{ir}}, \qquad \qquad \qquad \qquad \qquad \qquad \ \ \forall r \in \CurrentRoutes \label{eq:res_alloc} \\
    &(\allo_{ir} \wedge \allo_{ir'}), \implies   \nonumber \\
    & (\Start_r \geq \End_{r'} + C \cdot \length_r ) \vee (\Start_{r'} \geq \End_{r} + C \cdot \length_{r'}), \nonumber \\
    & \qquad \qquad \qquad \qquad \qquad \forall i \in V, \ r,r'\in \CurrentRoutes, \ r \neq r' \label{eq:non_overlap}  \\
    &\bigvee_{\allo_{ir} \in \lambda}{\neg{\allo_{ir}}} \qquad \qquad \qquad \qquad \qquad \qquad \   \forall \lambda \in \PreviousAssignment  \label{eq:previous_assignments}
\end{align}
Constraint~\eqref{eq:exactly_one_assignment} guarantees that exactly one vehicle is assigned to each route;
\eqref{eq:end_based_on_length} connects the \emph{start} and \emph{end} variables based on the route's length and their cumulative service time; \eqref{eq:latest_start} constrains the latest start time of a route to the strictest time window of its jobs;
\eqref{eq:res_alloc} expresses that one (or more) among the eligible vehicles must be assigned to a route; \eqref{eq:non_overlap} expresses that any two routes assigned to the same vehicle cannot overlap in time; one must end before the other starts. Finally, constraint \eqref{eq:previous_assignments} guarantees to find an assignment different from the already found ones.

Based on the model described above, the algorithm \emph{Assign} is defined, that takes the set of current routes \CurrentRoutes from the routing problem as input, and returns the current assignment \CurrentAssignment that specifies which vehicle that will use each route (and execute its jobs) and when it starts; if the \emph{Assignment Problem} is infeasible, $\CurrentAssignment = \emptyset$.  

\subsection{Capacity Verification Problem}

In this phase the goal is to find a feasible schedule for the vehicles, if it exists, meaning that the routes they are assigned to are evaluated to verify that capacity constraints are fulfilled. 
To do this, an ordered list of nodes $\NodeList_r$ and an ordered list of edges  $\EdgeList_r$, $\forall r \in \CurrentRoutes$, respectively, are generated, that each route visits.
Let $n_{\textrm{re}}$ be the node visited before edge $e$ on route $r$ and let $e_{\textrm{rn}}$ be the node visited before node $n$ on route $r$.
Similarly, let $n^{\textrm{re}}$ be the node visited after edge $e$ on route $r$ and let $e^{\textrm{rn}}$ be the node visited after node $n$ on route $r$.
Also, for each node in $\NodeList_r$ it is necessary to specify whether there exists a time window, since some of the nodes are only intersections of road segments in the real plant, while others are actual pickup or delivery points.
Let $l_{rn}$ and $u_{rn}$ be the earliest and latest arrival time, respectively, at node $n$ on route $r$;
let $\serviceTime_{rn}$ be the service time at node $n$ on route $r$, if such exists, \emph{zero} otherwise. 
Let $n^*_r$ be the starting node of route $r$.
Finally, let $e(1)$ and $e(2)$ be the source and sink node of edge $e$ respectively.

This phase is also treated as a JSP, where routes are jobs, while nodes and edges are the resources. Also, each route has a starting time $\Start_r$ defined by solving the \emph{Assignment Problem}. 
The decision variables in the \emph{Capacity Verification Problem} are:
\begin{itemize}[label={}]
    \item $\node_{rn}$: non-negative real variables that model when route $r$ is using node $n$;
    \item $\edge_{re}$: non-negative real variable that model when route $r$ is using edge $e$;
\end{itemize}
The model for the \emph{Capacity Verification Problem} is:
\begin{flalign}
    &\node_{rn^*_r} \geq start_r, \qquad \qquad \qquad \qquad \qquad \ \ \ \forall r \in \CurrentRoutes \label{eq:route_start} \\
    &\edge_{re} \geq \node_{rn_{re}} + \serviceTime_{rn_{re}}, \ \qquad \qquad \ \ \forall r \in \CurrentRoutes, \ e \in \EdgeList_r \label{eq:visit_precedence_1} \\
    &\node_{rn} = \edge_{re_{rn}} + d_{e_{rn}}, \ \ \ \qquad \qquad \forall r \in  \CurrentRoutes, \ n \in \NodeList_r \label{eq:visit_precedence_2} \\
    &\node_{rn} \geq l_{rn} \wedge \node_{rn} \leq  u_{rn}, \  \quad \qquad \ \forall r \in  \CurrentRoutes, \ n \in \NodeList_r \label{eq:visit_tw} \\
    &\node_{r_1n} \geq \edge_{r_2e^{r_1n}} + 1 \ \vee \ \node_{r_2n} \geq \edge_{r_1e^{r_2n}} + 1,  \nonumber \\
    & \qquad \qquad \qquad \qquad  \forall r_1,r_2 \in \CurrentRoutes, \  r_1  \neq r_2, \nonumber \\
    & \qquad \qquad \qquad \qquad \qquad  n \in \NodeList_{r1} \cap \NodeList_{r2}, \ n \notin \nodeset{H} \label{eq:nodes_no_swap}\\
    &\edge_{r_1e} \geq \edge_{r_2e} + 1  \vee   \edge_{r_2e} \geq \edge_{r_1e} + 1, \nonumber \\
    & \ \forall r_1,r_2 \in \CurrentRoutes, \, r_1  \neq r_2, \ e \in \EdgeList_{r_1} \cap \EdgeList_{r_2},\, g(e) = 1 \label{eq:edges_direct}\\
    &\edge_{r_1e_1} \geq \edge_{r_2e_2} +  d_{e_2} \ \vee \edge_{r_2e_2} \geq  \edge_{r_1e_1} + d_{e_1}, \nonumber \\
    & \quad \qquad \forall r_1,r_2 \in R, \ r_1 \neq r_2, \ e_1 \in \EdgeList_{r_1}, \nonumber \\
    & \qquad \qquad \qquad e_2 \in \EdgeList_{r_2}, \ e_1 = \bar{e}_2, \ g_{e_1} = g_{e_2} = 1  \label{eq:edges_inverse}
\end{flalign}

\noindent
\eqref{eq:route_start} constraints the start time of a route; \eqref{eq:visit_precedence_1} and \eqref{eq:visit_precedence_2} define the precedence among nodes and edges to visit in a route; \eqref{eq:visit_tw} enforces time windows on the nodes that correspond to the tasks; \eqref{eq:nodes_no_swap} prevents vehicles from using the same node at the same time (the $+1$ in the constraints forbids \emph{swapping} of positions between a node and the previous or following edge); \eqref{eq:edges_direct} and \eqref{eq:edges_inverse} constrain the transit of vehicles over the same edge. If two vehicles are using the same edge from the same node, one has to start at least one time-step later than the other and if two vehicles are using the same edge from opposite nodes, one has to be done transiting, before the other one can start. 

Based on the model described above, the algorithm \emph{CapacityVerifier} is defined, that takes the \CurrentAssignment from the \emph{Assignment Problem} as input and returns \CapacityVerifySolution, a list that expresses where each vehicle is at each time-step and, as for the previous phases, is empty if the problem is infeasible.

\subsection{Paths Changing Problem}

In this phase, alternative paths are computed to connect the consecutive tasks of each route. Finding alternative paths may be necessary when, for a given set of routes \CurrentRoutes, it is not possible to find any feasible schedule \CapacityVerifySolution. The infeasibility of the \emph{Capacity Verification Problem} may be due to the current set of paths \CurrentPaths that connect the tasks' locations, therefore a different set may lead to a feasible solution. We have previously defined a route as a sequence of tasks' locations and for any two consecutive tasks there is a path (a sequence of edges) connecting them. Therefore, for a route counting $i + 1$ tasks we will have $i$ paths and for each path we can define a start and an end node, respectively $\pathStart_i$ and $\pathEnd_i$. Finally, we define the sets of outgoing and incoming edges for a certain node $n$ as $\outgoingEdge_n$ and $\incomingEdge_n$, respectively.

Variables used to build the model are:
\begin{itemize}[label={}]
    \item $\useNode_{rin}$: Boolean variable that represents whether the $i$-th path of route $r$ is using node $n$;
    \item $\useEdge_{rie}$: Boolean variable that represents whether the $i$-th path of route $r$ is using edge $e$;
\end{itemize}

We could split this problem into $ r \cdot i$ problems (assuming all routes have $i + 1$ tasks) and find paths for each route separately; simpler models are faster. Unfortunately it may be necessary to explore different combinations of paths and so to retain the information we need we have only one model. Therefore, let the optimal solution to the \emph{Path Changing Problem} found at iteration \iteration be 
$$ \CurrentPaths = \bigcup_{ \substack{ r \in \CurrentRoutes \\ i= 1,\ldots,|r| \\ e \in \edgeSet } } { \{ \useEdge^*_{rie} \}, } $$
where $z^*_{rie}$, $\forall i = 1,\ldots,|r|, \ r \in \CurrentRoutes, \ e \in \edgeSet $, is the value of $z_{rie}$ in the current solution; also, let \PreviousPaths be the set containing the optimal solutions found until the $(\iteration-1)$-th iteration.
The model, similar to \cite{aloul2006identifying}, is as follows:
\begin{flalign}
    & \min_{i = 1,..,|r|, \ r \in \CurrentRoutes, \ n \in \nodeSet} \sum{\textrm{If}(\useNode_{rin},1,0)} \label{eq:path_changing_cost_function} \\
    & \useNode_{ri\pathStart_i} \wedge \useNode_{ri\pathEnd_i}, \qquad \qquad \quad \ \ \forall i = 1,\ldots,|r|, \ r \in \CurrentRoutes \label{eq:start_and_end_are_true} \\
    & \textrm{EN}_{e \in \outgoingEdge_{\pathStart_i} }{(\useEdge_{rie},1)}, \qquad \qquad \ \forall i = 1,\ldots,|r|, \ r \in \CurrentRoutes \label{eq:only_one_edge_for_start}\\
    & \textrm{EN}_{e \in \incomingEdge_{\pathStart_i} }{(\useEdge_{rie},1)}, \qquad \qquad \ \forall i = 1,\ldots,|r|, \ r \in \CurrentRoutes \label{eq:only_one_edge_for_end}\\
    & \useEdge_{rie} \implies \neg{\useEdge_{ri\bar{e}}}, \quad \ \ \forall i = 1,\ldots,|r|, \ r \in \CurrentRoutes, \ e \in \edgeSet \label{eq:not_both_directions}\\
    & \bigwedge_{n \in \nodeSet, n \neq \pathStart_i, n \neq \pathEnd_i}\textrm{If}(\useNode_{rin}, \nonumber \\
    & \qquad \qquad \textrm{EN}_{e \in \outgoingEdge_n}{(\useEdge_{rie}},1) \wedge \textrm{EN}_{e \in \incomingEdge_n}{(\useEdge_{rie},1)}, \nonumber \\
    & \qquad \qquad \qquad   \textrm{EN}_{e \in \outgoingEdge_n}{(\useEdge_{rie},0)} \wedge \textrm{EN}_{e \in             \incomingEdge_n}{(\useEdge_{rie},0)}), \qquad \nonumber \\
    & \qquad \qquad \qquad \qquad \qquad \quad \ \ \forall i = 1,\ldots,|r|, \ r \in \CurrentRoutes \label{eq:exactly_two_edges}\\
    &\bigvee_{\useEdge_{rie} \in \CurrentPaths}{\neg{\useEdge_{rie}}}, \qquad \qquad \qquad \qquad \qquad \quad \forall \CurrentPaths \in \PreviousPaths  \label{eq:previous_paths}
\end{flalign}

The cost function \eqref{eq:path_changing_cost_function} to minimize is the total number of used edges; \eqref{eq:start_and_end_are_true} guarantees that, for each path of each route, the start and end nodes are used; \eqref{eq:only_one_edge_for_start} and \eqref{eq:only_one_edge_for_end} make sure that exactly one outgoing (incoming) edge is incident with the start (end) node of a route; \eqref{eq:not_both_directions} makes sure that a path cannot use both an edge and its reverse; \eqref{eq:exactly_two_edges} guarantees that if a node (different from the start or end) is selected, exactly one of its outgoing and one of its incoming edges will be used. On the other hand, if a  node is not used, none of its incident edges will be used; finally, \eqref{eq:previous_paths} rules out all the previously found solutions.

Based on the model described above the algorithm \emph{Paths\-Changer} is defined, that takes the previous paths \PreviousPaths as input and returns
a new set of paths \newPaths, such that $\newPaths\cap\PreviousPaths=\emptyset$. If the \emph{Paths Changing Problem} is infeasible $\newPaths=\emptyset$.

\subsection{Routes Verification Problem}
The \emph{Routes Verification Problem} is a simplified version of the \emph{Routing Problem}, where a set of routes \CurrentRoutes already exists and it is verified whether these meet the requirements on the tasks' time windows and the vehicles' operating range. As described in Section~\ref{subsec:routing_problem}, routes are designed based (among other things) on the distance between tasks' locations; paths are computed between any two tasks' locations to have a uniquely defined distance and the routes designed accordingly in the \emph{Routing Problem}. However, as soon as the paths used to connect the tasks' locations are changed, there is no guarantee that the routes still meet the requirements, hence the need to verify the routes. 

Let $\taskSet_r = \{k_1,\ldots,k_{|r|} \} $ be the set of task for route $r \in \CurrentRoutes$, the variables used to build the model for the \emph{routes verification problem} are:

\begin{itemize}[label={}]
     \item $\sigma_{rk}$: non-negative real variable that models the time when task $k$ of route $r$ is served
    \item $\omega_{rk}$: non-negative real variable that models the remaining charge of a vehicle assigned to route $r$ when it reaches task $k$
\end{itemize}

The model is as follows:
\begin{flalign}
    & \omega_{rk} \cdot \rho \leq \OR, \qquad \qquad \qquad \qquad \forall k \in \taskSet_r,\, r \in \CurrentRoutes  \label{eq:checker_domain}\\
    & \sigma_{rk_{i+1}} \geq \sigma_{rk_i} + d_{k_ik_{i+1}} + \serviceTime_{k_i},  \nonumber \\
    & \qquad \qquad \quad \ \ \qquad \qquad \qquad \forall i = 1,\ldots,|r|, \, r \in \CurrentRoutes \label{eq:checker_arrival_time} \\
    & \sigma_{rk} \geq l_{k} \wedge \sigma_{rk} \leq u_{k}, \qquad \qquad \quad \ \forall k \in \taskSet_{j}, \, j \in \jobSet \label{eq:checker_time_window} \\
    & \omega_{rk_{i+1}} \leq \omega_{k_i} - D \cdot d_{k_{i}k_{i+1}}, \nonumber \\
    & \qquad \qquad \qquad \qquad \qquad \quad \ \ \forall i = 1,\ldots,|r|, \, r \in \CurrentRoutes \label{eq:checker_autonomy} 
\end{flalign}
\noindent
\eqref{eq:checker_domain} restricts the domain of the remaining charge to be smaller than or equal to the operating range of the vehicles; \eqref{eq:checker_arrival_time} connects the arrival time at the task based on the distance between them;
\eqref{eq:checker_time_window} forces the arrival time at a task's location to be within its time window; \eqref{eq:checker_autonomy} relates the remaining charge when reaching a task's location to the distance from the previous task's location.

Based on the model described above the algorithm \emph{RoutesVerifier} is defined, that takes the current routes \CurrentRoutes and the current paths \CurrentPaths and returns  true if the problem is feasible, false otherwise.

\subsection{Solving the \texorpdfstring{\theModel}{theModel} using the \texorpdfstring{\theAlgorithm}{theAlgorithm} algorithm}

\begin{figure}
    \centering
    \includegraphics[width = 0.4\textwidth]{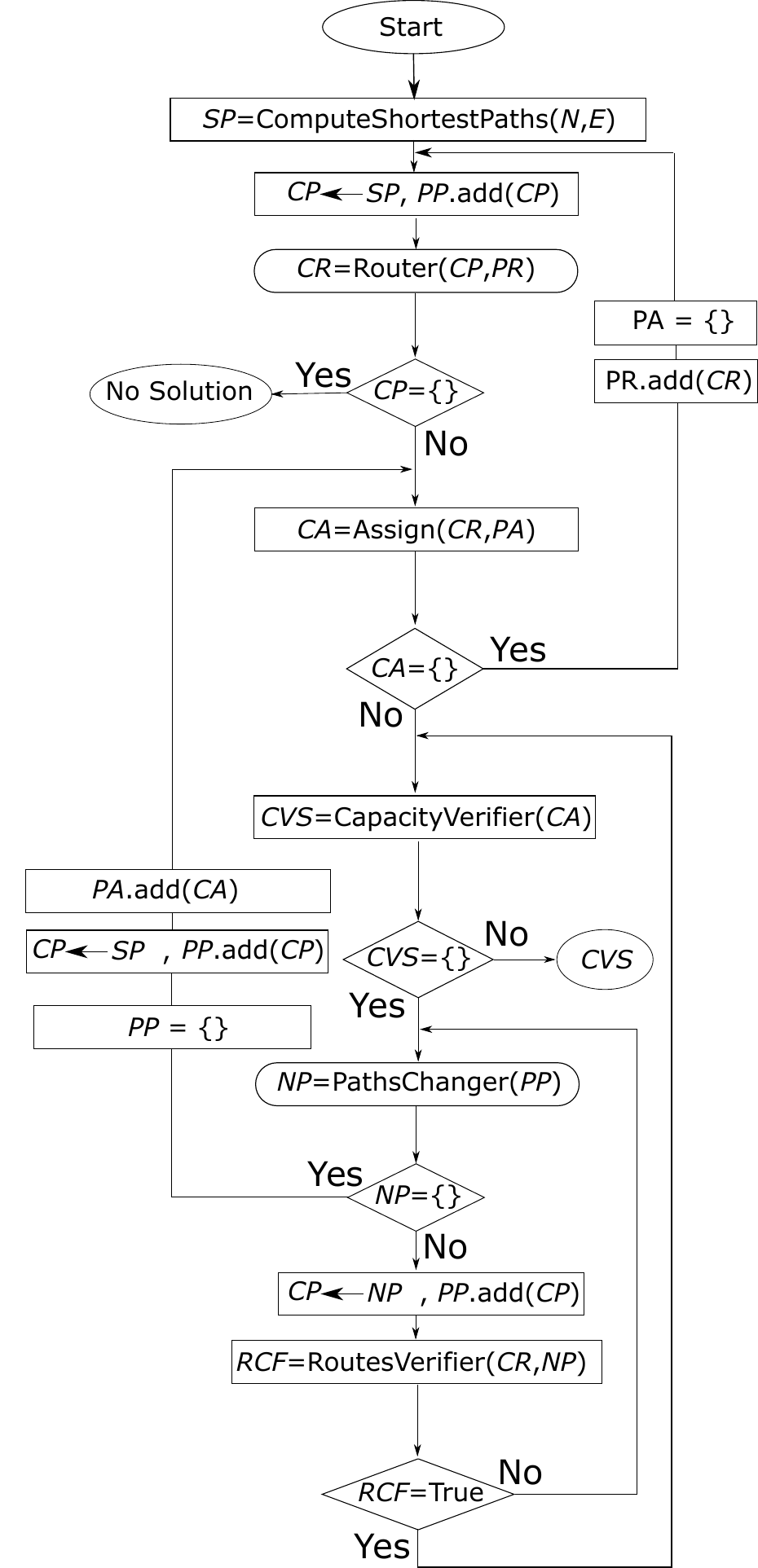}
    \caption{Flowchart of \theAlgorithm algorithm.}
    \label{fig:flowchart}
\end{figure}

The \emph{\theAlgorithm} algorithm, Figure~\ref{fig:flowchart}, connects the above described sub-problems to find a feasible solution to the full problem. The \emph{Router} and \emph{PathsChanger} algorithms are in Figure~\ref{fig:flowchart} put in rounded corner boxes, to show that they are optimization problems.

The algorithm begins with the computation of the shortest paths between each pair of tasks. This step is only executed once to provide unique paths for the \emph{Routing Problem}, which is then solved. In this step neither the vehicles' availability nor the segment capacities are considered; the goal is simply to design routes to serve tasks within the time windows. Therefore, if the \emph{Routing Problem} is infeasible, the whole problem is infeasible, because there is no possible routing such that tasks are served within their time windows. The information about the previous routes will be stored so that each time this algorithm is called, it will provide a new solution to the \emph{Routing Problem}. 

If the \emph{Routing Problem} is feasible the next step is to verify whether the available vehicles can execute the routes. This matching is based on the routes' requirements for specific types of vehicles, on their latest start time, and on the vehicles' operating range and charge rate. This is done by solving the \emph{Assignment Problem}; also in this case there can be more feasible solutions, therefore it is important to store the current one to be able to rule it out the next time the \emph{Assignment Problem} is solved. 
If the \emph{Assignment Problem} is infeasible the algorithm backtracks and runs the \emph{Routing Problem} again, otherwise, it proceeds to the \emph{Capacity Verification Problem}.

At this point, routes have been assigned an actual vehicle to execute them and start times have been restricted to meet the vehicles' need for charging. Hence it is possible to verify if the execution of the routes is possible without breaking the capacity constraints. If that is the case, the overall problem is feasible and the algorithm terminates and returns a feasible schedule. On the other hand, if this step is infeasible, the algorithm will try to find alternative paths for the vehicles to execute the routes. 

This step is split in two parts. The \emph{PathsChanger} algorithm finds new paths and the \emph{RoutesVerifier} makes sure the \emph{Routing Problem} is still solvable (i.e. tasks can still be served within time windows) using these new paths. 
If the \emph{Paths Changing Problem} is infeasible, all paths from one task to the following one have been checked for each route.
Therefore the algorithm backtracks and looks for a assignment. Otherwise if the \emph{Paths Changing Problem} is feasible, the algorithm moves forward to the \emph{Routes Verification Problem}. If this problem is feasible the algorithm backtracks to verify whether it is feasible against the capacity constraint by solving the \emph{Capacity Verification Problem}; if not, the \emph{PathsChanger} algorithm is called again.

Whenever the \emph{Assignment Problem} is infeasible, all possible assignments for the current set of routes \CurrentRoutes have been explored. Thus, before calling the  \emph{Router} algorithm again, \CurrentRoutes is added to \PreviousRoutes. In the same way, whenever the \emph{Paths Changing Problem} is infeasible,  all possible paths for the current assignment \CurrentAssignment have been explored, hence \CurrentAssignment is added to \PreviousAssignment. Also, the set of previous paths \PreviousPaths is emptied because these paths are only eligible for the current assignment, and the shortest paths are set as current paths to compute the next assignment. 

\begin{table}[h] \label{fig:glossary}
    \normalsize
  \centering
  \caption{Glossary for the sets of the sub-problems.}
    \begin{tabular}{p{0.45\textwidth}}
    \toprule
    \CurrentPaths: set of current paths \\
    \ShortestPaths: set of shortest paths \\
    \newPaths: set of new paths \\
    \PreviousPaths: set of previous paths \\
    \CurrentRoutes: set of current routes \\
    \PreviousRoutes: set of previous routes \\
    \CurrentAssignment: set of current assignment of vehicles to routes \\
    \PreviousAssignment: set of previous assignment of vehicles to routes\\
    \CapacityVerifySolution: set of \emph{conflict-free} routes (pairs of nodes and arrival times for each route) \\
    \RoutesVerifyingFeasibility: Boolean variable representing the feasibility of the \emph{Routing Problem} \\
    \bottomrule
    \end{tabular}%
  \label{tab:glossary}%
\end{table}%

On the other hand, when exploring different paths  the current assignment is not changed, it is only checked whether it is feasible with the new paths; thus, \CurrentAssignment is not added to \PreviousAssignment. Finally, as \theAlgorithm loops through \emph{PathsChanger} and \emph{Routes\-Verifier} to find a feasible set of paths, \newPaths is assigned to \CurrentPaths, which in turn is added to \PreviousPaths after every unsuccessful iteration.

The glossary of Table~\ref{tab:glossary} contains the names of the sets used to store and exchange information among the sub-problems presented in Section~\ref{sec:compo_algo}.

\subsection{Solving the Example using \theAlgorithm}

\begin{figure}
    \centering
    \includegraphics[width = 0.3\textwidth]{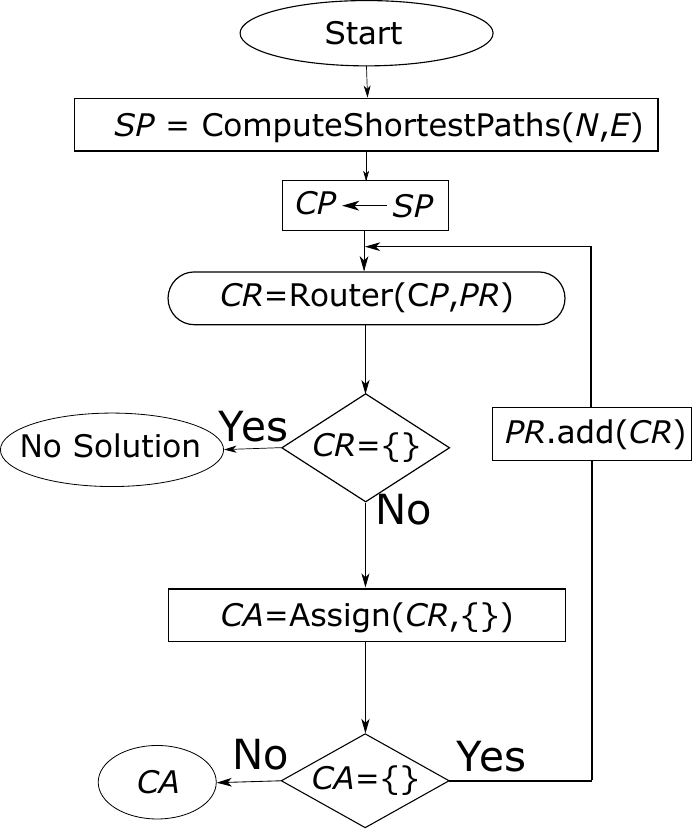}
    \caption{Flowchart of \theRelaxedAlgorithm.}
    \label{fig:relaxed_problem}
\end{figure}

To illustrate the performance of \theAlgorithm the example of Section~\ref{sec: problem_definiton} is used.
Below are reported the function calls to solve the sub-problems, whether they are feasible or infeasible, and their running times:
\begin{itemize}
    \item[] \emph{Router}: \emph{feasible}, 0.45\,s
    \item[] \emph{Assign}: \emph{feasible}, 0.05\,s
    \item[] \emph{CapacityVerifier}: \emph{feasible}, 0.08\,s
\end{itemize}

The implementation of \theAlgorithm has the ability of manually setting the solution to a sub-problem to be \emph{infeasible}. We set the solution to the \emph{Capacity Verification Problem} to be \emph{infeasible} to evaluate the performance of the algorithm when conflicts arise due to capacity constraints. We also set a limit of 50 on the number of alternative sets of paths to generate. Of these 50 sets, 29 were declared infeasible by the \emph{RoutesVerifier}. The average running time to solve the \emph{Path Changing Problem} is 0.7\,s; as for the \emph{Routes Verification Problem}, it took on average 0.01\,s to solve it, regardless of the feasibility of the sub-problem.

For the larger problem instances discussed in Section~\ref{sec:experiments} counting up to 11 vehicles and 15 jobs, the \emph{Router} and the \emph{PathChanger} calls take roughly 10\,s each, while the \emph{CapacityVerifier} takes about 1.5\,s, \emph{Assign} takes less than 0.5\,s, and \emph{RoutesVerifier} takes between 0.01\,s and 0.1\,s.

\section{Soundness and Completeness of \texorpdfstring{\theAlgorithm}{theAlgorithm}} \label{sec: proofs}

This section provides proof of \theAlgorithm's soundness and completeness. We start by relaxing the capacity constraints on the segments. We refer to the relaxed version of \theAlgorithm as the \emph{Capacity Relaxed-\theAlgorithm} (\emph{\theRelaxedAlgorithm}). This turns the problem into a combination of routing and assignment. We use the conclusions from \theRelaxedAlgorithm as a starting point to prove the soundness and completeness of \theAlgorithm.

When we do not have to deal with capacity constraints we can simplify \theAlgorithm, as shown in Fig.~\ref{fig:relaxed_problem}; we essentially have to solve the \emph{Routing Problem} and then verify that the solution found has a feasible assignment by solving the \emph{Assignment Problem}. If that is the case, the algorithm terminates and returns a feasible solution, else it tries to design different routes. If no routing solution has a feasible assignment, the algorithm terminates with \emph{No Solution}.

\begin{obs} \label{obs: all_problems_decidable}
    All the problems solved in \theAlgorithm are decidable. This is true because they are all combinations of decidable first-order theories and therefore the \emph{Nelson-Oppen theory combination method}~\cite{tinelli1996new} applies. In fact the \emph{Routing Problem} is a combination of linear arithmetic and propositional logic, the \emph{Assignment Problem}, \emph{Capacity Verification Problem}, and \emph{Routes Verification Problem} all fall into the category of difference logic (a fragment of linear arithmetic), and the \emph{Paths Changing Problem} is a propositional logic problem. 
\end{obs}

\begin{obs} \label{obs:problem_is_bounded}
    The optimization problems solved in \theAlgorithm, i.e., the \emph{Routing Problem} and the \emph{Paths Changing Problem}, are bounded. The \emph{Routing Problem} involves a finite number of decision variables that are either Booleans with a finite domain, or non-negative integers and the objective is minimization. The \emph{Paths Changing Problem} involves only Boolean variables, so the domain is finite.
\end{obs}

\begin{lemma} \label{lemma: shortest_paths}
    Given a problem instance of the \theModel, if a feasible solution to the \emph{Routing Problem} cannot be found using the shortest paths to connect any two tasks, no feasible solution can be found using any other set of paths.
\end{lemma}

\begin{proof}
    The \emph{Routing Problem} can be infeasible for two reasons (or a combination of them): there exist no routes such that all tasks' time windows can be met; there exist no routes shorter than or equal to the vehicles operating range such that all tasks are served. 
    
    When it comes to the time windows, the lower bound does not affect the feasibility of an instance, because vehicles are allowed to wait at the task's location before starting the service. On the other hand, if there is no way a vehicle can reach a task's location before the time window's upper bound the instance is infeasible. Since it is assumed that vehicles travel at constant speed, the distances among tasks' locations are directly proportional to the time required to travel between them; therefore, if time windows cannot be met travelling along the shortest paths, neither can they using any other set of paths.
    
    As for the routes maximum length, restricted by the vehicles operating range, the same reasoning applies. If it is not possible to design routes to serve all tasks that are shorter than or equal to the vehicles operating range using the shortest paths, neither will it be using longer paths.
\end{proof}

Lemma~\ref{lemma: shortest_paths} is required both for \theRelaxedAlgorithm and \theAlgorithm. When relaxing the capacity constraints, 
before ruling out a set of routes it needs to be made sure that they are infeasible.
If arbitrarily long paths were used, there would still be a chance that using other paths could make the \emph{Routing Problem} feasible. As for \theAlgorithm, if a set of routes is infeasible using the shortest paths, there is no need to try to replace paths and check for feasibility.

\subsection{Relaxed Problem: Capacity constraints not included}

When relaxing the capacity constraints the problem boils down to designing the routes, which is taken care of by the \emph{Routing Problem}, and the assignment of the available vehicles, handled by solving the \emph{Assignment Problem}. 
\begin{lemma} \label{lemma:routing_finite}
    The \emph{Routing Problem} has a finite number of feasible solutions. 
\end{lemma}

\begin{proof}
    Let $\varphi$ be the conjunction of constraints \eqref{eq:not_travel_same_spot}-\eqref{eq:routing_precedence}, let $\varphi^r$ be the conjunction of constraints \eqref{eq:not_travel_same_spot}-\eqref{eq:route_continuity}, and let $\varphi^s$ be the conjunction of constraints \eqref{eq:routing_time_window}-\eqref{eq:routing_precedence}. $\varphi^r$ does not take into account tasks, jobs, operating range, or time windows; it only guarantees that routes are closed and that each task is visited exactly once.
    Let $k = |\taskSet|$ and let a route $r$ be an ordered subset of \taskSet. 
    Then, a solution to the routing problem $R = \{\,r_1,\ldots,r_m \,|\, m \leq k \}$ is a partition of the set \taskSet. The number of partitions of the set \taskSet is $ \sum_{q=0}^k{\binom{k}{q}} < \infty,$ which corresponds to the number of possible solutions for $\varphi^r$, $\mathbb{S}(\varphi^r)$. Since $\varphi = \varphi^r \wedge     \varphi^s$, then $\mathbb{S}(\varphi) \leq \mathbb{S}(\varphi^r) < \infty$    
\end{proof}

\begin{lemma} \label{lemma:enumerate_all_routes}
    Repeated calls to \emph{Router} will enumerate all feasible solutions before returning infeasible.
\end{lemma}

\begin{proof}
    Let $\varphi_0$ be a relaxation of the \emph{Routing Problem}, not including constraint \eqref{eq:routing_precedence}, and let $\CurrentRoutes_0$ be a solution to $\varphi_0$. Then, if another solution $\CurrentRoutes_1$ for $\varphi_0$ exists, it can be found by solving $\varphi_0 \wedge \neg{\CurrentRoutes_0} = \varphi_1$. In general, the $n$-th solution can be found by solving $\varphi_0 \wedge \neg{\CurrentRoutes_0}\wedge\ldots\wedge\neg{\CurrentRoutes_{n-1}} = \varphi_{n}$. Because of Lemma~\ref{lemma:routing_finite}, we know that $\mathbb{S}(\varphi) < \infty$ and we  enumerate  all by solving $\varphi_0,\,\ldots,\,\varphi_{\mathbb{S}(\varphi)-1}$
\end{proof}

\begin{theorem} \label{theorem:simplified_algo_is_complete}
    \theRelaxedAlgorithm in Fig.~\ref{fig:relaxed_problem} is sound and complete.
\end{theorem}

\begin{proof}
    For a problem where $|\taskSet|$ tasks are to be executed and $|\vehicleSet|$ vehicles are available, a solution is an assignment $\CurrentAssignment = \{(v,r_1)_1,\ldots,(v,r_i)_i \} \ v \in \vehicleSet, \ i \leq |\taskSet| $ that satisfies the \emph{Assignment Problem} and $\bigcup_{i \leq |\taskSet|}{r_i}$ is a feasible solution to the \emph{Routing Problem}. Because of \emph{Lemma}~\ref{lemma:routing_finite} we know there is only a finite number of solutions to the \emph{Routing Problem}, and because of \emph{Lemma}~\ref{lemma:enumerate_all_routes} we know we can enumerate them all. In the algorithm, for each solution \CurrentRoutes we check whether it satisfies the \emph{Assignment Problem}; hence, if the overall problem has a feasible solution, the algorithm will eventually find it, otherwise it will declare the problem infeasible.
\end{proof}

\subsection{Full Problem}

When including the capacity constraints, the paths chosen to move from one task's location to another become crucial. For \theAlgorithm to be sound and complete, we need to prove that it can explore all possible paths for a given set of routes and a given assignment. Note that we have restricted the problem to forbid cycles in the paths; however, in some instances of the \theModel cycles may be required. We are going to show that, if we use a different model to change paths such that also cycles are allowed, the algorithm is sound and complete for any instance of the \theModel. 

\begin{lemma} \label{obs:finite_number_of_paths}
    Given a directed, weighted graph with a finite number of nodes, the number of paths that connect two arbitrary nodes is finite.
\end{lemma}

\begin{proof}
    By definition, a path is a sequence of edges that joins a sequence of nodes and no node appears more than once. If the number of nodes in the graph is finite, there cannot be an infinite number of sequences of nodes to connect to arbitrary nodes. 
\end{proof}

\begin{lemma} \label{lemma:enumerate_all_paths}
    For a given set of routes \CurrentRoutes and a given assignment of vehicles \CurrentAssignment, repeated calls to the \emph{Paths\-Changer} algorithm will enumerate all feasible solutions before returning infeasible. 
\end{lemma}

\begin{proof}
    Let $\varphi_0$ be the conjunction of constraints \eqref{eq:start_and_end_are_true}-\eqref{eq:exactly_two_edges}, a relaxation of the \emph{Paths Changing Problem}, and let $\CurrentPaths_0$ be a solution to $\varphi_0$. Then, if another solution $\CurrentPaths_1$ for $\varphi_0$ exists, it can be found by solving $\varphi_0 \wedge \neg{\CurrentPaths_0} = \varphi_1$. In general, the $n$-th solution can be found by solving $\varphi_0 \wedge \neg{\CurrentPaths_0}\wedge\ldots\wedge\neg{\CurrentPaths_{n-1}} = \varphi_{n}$. Because of \emph{Lemma}~\ref{obs:finite_number_of_paths}, we know that the number of solutions to the \emph{Paths Changing Problem} $\mathbb{S}(\varphi) < \infty$ and we enumerate all by solving $\varphi_0,\,\ldots,\,\varphi_{\mathbb{S}(\varphi)-1}$.
\end{proof}

When considering the capacity constraints, different assignments may lead to different solutions; for instance one vehicle may not be available until a certain time because it is still recharging after executing a route while another is available earlier. Both assignments are feasible but they will execute the routes at different time, hence the road segments will be occupied by the vehicles at different times, which in turn may lead to different schedules. Therefore it is necessary to explore all possible assignments. 

\begin{lemma} \label{lemma:assignment_finite}
    The \emph{Assignment Problem} has a finite number of feasible solutions. 
\end{lemma}

\begin{proof}
      In an instance of the \theModel we assume to have a finite number of vehicles $v = |\vehicleSet|$. We know from Lemma~\ref{lemma:routing_finite} that, given a finite number of tasks, a solution to the \emph{Routing Problem} has at most as many routes as tasks. Let $\varphi$ be the conjunction of constraints \eqref{eq:exactly_one_assignment}-\eqref{eq:non_overlap}, let $\varphi^r$ be the conjunction of constraint \eqref{eq:exactly_one_assignment}, and let $\varphi^s$ be the conjunction of constraints \eqref{eq:end_based_on_length}-\eqref{eq:non_overlap}. $\varphi^r$ is a relaxation of the \emph{Assignment Problem} that does not take into account routes length, charging time or vehicle eligibility, it only guarantees that exactly one vehicle is assigned to each route. A solution to the \emph{Assignment Problem}, represented by $\varphi^r$, is therefore a partition of the routes set \CurrentRoutes. Let $c = |\CurrentRoutes|$, then the number of partitions of the set \CurrentRoutes is $ \sum_{q=0}^c{\binom{c}{q}} < \infty,$ which corresponds to the number of possible solutions for $\varphi^r$, $\mathbb{S}(\varphi^r)$. Since $\varphi = \varphi^r \wedge \varphi^s$, then  $\mathbb{S}(\varphi) \leq \mathbb{S}(\varphi^r) < \infty$ 
\end{proof}

\begin{lemma} \label{lemma:enumerate_all_assignments}
     For a given set of routes \CurrentRoutes, repeated calls to \emph{Assign} will enumerate all feasible solutions before returning infeasible. 
\end{lemma}

\begin{proof}
    Let $\varphi_0$ be the conjunction of constraints \eqref{eq:end_based_on_length}-\eqref{eq:non_overlap}, a relaxation of the \emph{Assignment Problem}, and let $\CurrentAssignment_0$ be a solution to $\varphi_0$. Then, if another solution $\CurrentAssignment_1$ for $\varphi_0$ exists, it can be found by solving $\varphi_0 \wedge \neg{\CurrentPaths_0} = \varphi_1$. In general, the $n$-th solution can be found by solving $\varphi_0 \wedge \neg{\CurrentAssignment_0}\wedge\ldots\wedge\neg{\CurrentAssignment_{n-1}} = \varphi_{n}$. Because of \emph{Observation}~\ref{lemma:assignment_finite}, we know that the number of solutions to the \emph{Assignment Problem} $\mathbb{S}(\varphi) < \infty$ and we enumerate all by solving $\varphi_0,\,\ldots,\,\varphi_{\mathbb{S}(\varphi)-1}$.
\end{proof}

\begin{theorem}
    \theAlgorithm is sound and complete.
\end{theorem}

\begin{proof}
    For a problem with $|\taskSet|$ tasks, $|\vehicleSet|$ vehicles, and a graph 
    $G(\nodeSet,\edgeSet)$, let $\overline{r}$ be the sequence of nodes visited to execute route $r$ and $\tau_{rn}$ the arrival time at node $n$ of route $r$; a solution is a schedule for each vehicle $v \in \vehicleSet$, 
    \begin{flalign}
    \CapacityVerifySolution = \{&(v,((n_{11},\tau_{11}),\ldots,(n_{1\overline{r}_1},\tau_{1\overline{r}_1})))_1,\ldots, \nonumber \\
    &(v,((n_{i1},\tau_{i1}),\ldots,(n_{i\overline{r}_i},\tau_{i\overline{r}_i})))_i \}, \,\forall i \leq |\taskSet|,  \nonumber
    \end{flalign}   
    that satisfies the \emph{Capacity Verification Problem}. From \emph{Lemma}~\ref{lemma:enumerate_all_routes} we know that we can enumerate all possible routes and from \emph{Lemma}~\ref{lemma:enumerate_all_assignments} we know that, for each set of routes we can enumerate all assignments. For each assignment $(v,r)$ the arrival time of vehicle $v$ at a node depends on the paths chosen to travel from one task of route $r$ to the following one. Since we know from \emph{Lemma}~\ref{lemma:enumerate_all_paths} that for a current set of routes \CurrentRoutes, and an assignment \CurrentAssignment of vehicles to it, we can enumerate all paths from one task of each route to the following one, if there exists a solution to the problem, \theAlgorithm will eventually find it; otherwise it will correctly declare the problem infeasible.
\end{proof}

As mentioned before, the \emph{PathsChanger} can only return (non-cyclic) paths. We know that there is a finite number of paths in a graph to go from one node to another, this is not true if cycles are allowed. On the other hand, even if cycles were allowed, if we limited the paths' maximum length, we could enumerate them all. Since we have time windows on the tasks and a limited operating range for the vehicles, we can compute an upper bound for the length of the pahts.

Therefore, if we were to modify the \emph{PathsChanger} to allow for cycles, the algorithm would still be complete without restricting the problem to non-cyclic paths. This feature is currently under investigation as future work.

\section{Evaluation} \label{sec:experiments}

In order to evaluate \theAlgorithm, a set of benchmark problems is proposed. The parameters for generating the benchmark problems are the number of nodes, vehicles, and jobs (grouped into the parameter N-V-J), as well as the time horizon, and the \emph{edge reduction} value, which is inversely proportional to the connectivity of the graph (the higher the value, the fewer edges). Vehicles can be of type \emph{A}, \emph{B}, or \emph{C}, and jobs come with a set of types that are eligible to execute them. 
For each combination of these parameters, five different problems were randomly generated. Problems belonging to the same category differ from each other in terms of tasks locations (including the additional tasks representing the depots), service time, time window (generated as a function of the time horizon), and vehicles eligible to execute them; other parameters that differ within the same category are the vehicles' operating range, the charging coefficient, and the number of vehicles available per type. 
Both \MonoMod (see below) and the algorithms called by \theAlgorithm used Z3 4.8.9 to solve the models. All the
experiments\footnote{The implementation of the algorithm presented in Section~\ref{sec: problem_definiton} and the problem instances are available at \url{https://github.com/sabinoroselli/VRP.git}.}
were performed on an \emph{Intel Core i7 6700K, 4.0 GHZ, 32GB RAM}  running \emph{Ubuntu-18.04 LTS}.

As mentioned in Section~\ref{sec: literature_review}, to the best of our knowledge, the \theModel presented in ~\cite{roselli2021smt} and further developed in this work is novel. 
The experimental evaluation compares the monolithic model, \MonoMod, presented in~\cite{roselli2021smt} against \theAlgorithm. Both the running time and quality of solutions are evaluated with respect to the problem parameters.

The first set of experiments compare the monolithic model (\MonoMod) of the \theModel  presented in~\cite{roselli2021smt} , to \theAlgorithm on a set of relatively small problems, with up to four vehicles, seven jobs, and a time horizon of 60 time-steps; the time limit set for both methods was 1200\,s. For this comparison \MonoMod has been adapted to account for non-negligible service times and the the cost function is not included in the model, so that \MonoMod returns the first feasible solution found.
The choice of 1200\,s is motivated by the industrial application the algorithm is designed for; while some schedules may be computed hours before they actually take place, last minute changes may happen and it is useful to know what size of problems can be solved within minutes.
For \MonoMod the model generation time may not be negligible; however, the comparison with \theAlgorithm is for the solving time. 

Table~\ref{tab:comparison_table} shows the results of the comparison. For smaller problems and a small time horizon, \MonoMod is performing well, often outperforming \theAlgorithm, especially when the problems are infeasible. As the problems grow larger though, \theAlgorithm performs better both in terms of solving time, and in terms of number of problems solved within the time limit. As expected, a larger time horizon has a negative impact on the solving time of \MonoMod, since the model is based on time discretization, and a larger \emph{T} means more variables and more constraints. On the other hand, the time horizon does not seem to affect the performance of \theAlgorithm significantly; instances having the same value of N-V-J and \emph{edge reduction}, and increasing time horizon show similar solving time. There are exceptions, but they may also be due to the different time windows, since these are generated based on the time horizon. The increase of the parameter \emph{edge reduction} generally corresponds to an increase in the solving time for both \MonoMod and \theAlgorithm, probably because having fewer edges makes it harder to find a solution if it exists, or prove infeasibility otherwise, though there are exceptions. Finally, the increase of the N-V-J parameter, as expected, corresponds to longer solving time in most cases. The reason behind the outliers, i.e. when a problem is immediately declared feasible/infeasible, is often the triviality of the problem itself. For example, when deadlines are too strict and there is no solution to the \emph{Routing Problem} then \theAlgorithm will terminate early.

The second set of experiments evaluate the performance of \theAlgorithm on a set of larger problems, with up to eleven vehicles, fifteen jobs and a time horizon of 300 time units. Again the time limit was set to 1200\,s. As for the previous set of instances, the increase in the value of N-V-J corresponds to an increase in the average solving time and a decrease in the number of solved instances per category. This time, infeasible instances are generally easier to solve, probably because of the higher number of jobs compared to the number of available vehicles (trivial infeasibility). 

Overall, the evaluation showed that \theAlgorithm's performance highly depends on the problem instance; there have been rather small instances that took a long time to solve, while other relatively large instances were solved almost immediately. In general, infeasibility seems to be harder to show than feasibility. This behaviour does not come unexpected, since for \theAlgorithm, a problem cannot be declared infeasible till all solutions have been explored. Moreover, as the number of \PreviousRoutes stored grows, finding a new solution becomes harder. For some problems, infeasibility  may  be  trivial  to  prove,  when  the operating  range  is  not  large  enough  or  the  time  windows are too strict, for instance. In other cases it may take several attempts before declaring a problem infeasible.

As for feasible problems, a similar reasoning applies. Sometimes it took many attempts to find a set of routes that actually led to a feasible schedule and, in general, the likelihood of finding one decreases as the number of vehicles and jobs increases. Nevertheless, even for large problems, a solution can be found rather quickly, given that enough vehicles are available.

\subsection*{Discussion on Optimality}

So far, the focus of the experiments was on the running time required by \MonoMod and \theAlgorithm to solve instances of the \theModel but no on the quality of the solutions. This section focuses on the quality of the solutions produced by \theAlgorithm for a set of problem instances by comparing them to a lower bound manually computed by relaxing the capacity constraints. \MonoMod has been set up to find optimal solutions by including a cost function representing the total travelled distance. To make the comparison possible, the parameters of the problem instances have been scaled down to make the problems simple enough so that
the optimal solution can be found by \MonoMod in reasonable time. Table~\ref{tab:goodness_comparison} shows the running time and cost function value for a set of \theModel instances. The problems are sorted by size, in terms of the parameters previously discussed\footnote{Details of the problem instances are available at \url{https://github.com/sabinoroselli/VRP.git} in
the file \emph{Optimality\_test\_instances.pdf}.}.
For the instances 1 to 6, both \theAlgorithm and \MonoMod have the same total travelled distance. Hence, for these instances, \theAlgorithm indeed returns the optimal solution. 
For the instances 7 to 9, \MonoMod was not able to return a solution after 24 hours and was therefore timed out. However, the cost function value returned by \theAlgorithm matches the lower bound, hence the solutions are optimal. 
For the instances 1 to 3, both \theAlgorithm and \MonoMod return a value higher than the lower bound, implying that the capacity constraints forced the vehicles to travel through paths longer than the shortest ones in order to serve the customers. 

\begin{table}[htbp]
  \centering
  \caption{Comparison of the Cost Function Value (CFV), and running time (in seconds) required to solve problem instances of the \theModel. For each instance, a Lower Bound (LB) on the cost function is also provided. The time limit is set to 24 hours and ``-'' means that this limit was exceeded.}
  
    \begin{tabular}{cccccc}
    \toprule
    \multirow{2}[2]{*}{Instance} & \multirow{2}[2]{*}{LB} & \multicolumn{2}{c}{ComSat} & \multicolumn{2}{c}{MonoMod} \\
       &    & CFV & Time & CFV & Time \\
    \midrule
    1  & 10 & 12 & 0.13 & 12 & 0.40 \\
    2  & 18 & 22 & 10.72 & 22 & 1.21 \\
    3  & 26 & 30 & 45.40 & 30 & 2.30 \\
    4  & 20 & 20 & 0.53 & 20 & 3.24 \\
    5  & 48 & 48 & 0.30 & 48 & 4.00 \\
    6  & 48 & 48 & 0.44 & 48 & 4.31 \\
    7  & 62 & 62 & 0.79 & -  & - \\
    8  & 72 & 72 & 0.93 & -  & - \\
    9  & 76 & 76 & 0.95 & -  & - \\
    10 & 10 & 12 & 0.15 & 10 & 0.19 \\
    \bottomrule
    \end{tabular}%
  \label{tab:goodness_comparison}%
\end{table}%

\begin{figure}[h]
    \centering
    \includegraphics[width = 0.3\textwidth]{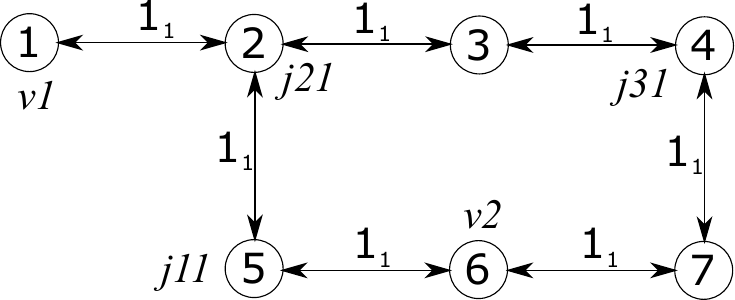}
    \caption{Finite, strongly connected, weighted, directed graph representing the plant layout for a problem instance of the \theModel that cannot be solved to optimality by \theAlgorithm.}
    \label{fig:counter_example_optimality}
\end{figure}

However, \theAlgorithm is not guaranteed to find the optimal solution, as shown by instance 10 above. This is due to the way the sub-problems are structured. For a given set of routes \theAlgorithm will try to find a feasible set of paths that satisfies the capacity constraints. If such set of paths exists, \theAlgorithm terminates with a feasible solution. Nevertheless, there could exist another set of routes for which there exists a set of paths that are cumulatively shorter and satisfies the capacity constraints. 
This is clarified with the following example, depcited in Fig.~\ref{fig:counter_example_optimality}.

\begin{itemize}[label={},leftmargin=*]
    \item $ \nodeSet = \{ 1,\ldots,7 \}, \ \nodeset{H} = \emptyset,  \Origin = \{ 1,6 \}$
    \item $ \edgeSet = \{(1,2),(2,3),(2,5),(3,4),(4,7),(5,6),(6,7)\} $
    \item $\jobSet = \{j1,j2 \}, \ \taskSet = \{i1,i2 \ | \ \forall i \in \jobSet  \}$
    \item $ L_{j11} = 5, \ L_{j21} = 2, \ L_{j31} = 4$
    \item $ \precTask_{j11} = \emptyset, \precTask_{j21} = \emptyset, \precTask_{j31} = \emptyset $
    \item $ l_{j11} = 2, \ l_{j21} = 2, \ l_{j31}= 2 $
    \item $ u_{j11} = 2, \ u_{j21} = 5, \ u_{j31}= 7 $
    \item $\serviceTime_{j11} = 2, \ \serviceTime_{j21} = 1, \serviceTime_{j31} = 1 $
    \item $\vehicleSet = \{ v1,v2 \}, \ \vehicleSet_{j1} = \{ v1 \}, \ \vehicleSet_{j2} = \{ v2 \}, \ \vehicleSet_{j3} = \{ v2 \} $
    \item $ \OR = 10, \C = 1, \ \D = 1, \ \rho = 1, \ \speed = 1, \ \timeHorizon = 13 $
\end{itemize}

Vehicle $v1$ needs to travel from node\,1 to node\,5 in order to execute task $j11$; vehicle $v2$ will be assigned to both task $j21$ and $j31$, respectively located at nodes 2 and 4. Since these tasks are equidistant from $v2$'s location (node\,6), serving one before the other or the other way around would result in the same cost for a route, hence \theAlgorithm may choose one as well as the other. Assuming that task\,$j21$ is served before $j31$, there will be a conflict; in fact, vehicle\,$v1$ occupies node\,5 to serve task\,$j11$, due to its time window and service time. If \theAlgorithm had come to such a situation, it would call the \emph{PathChanger} function. There is no different path for $v1$ that would be feasible against the time window of task\,$j11$; however, $v2$ could reach node\,2 by passing through nodes 7, 4, and 3 and still meet the time window of task\,$j21$. It would then go back to node\,4 and execute task\,$j31$. The total length of such a route will be 8. However, serving task\,$j31$ before $j21$ would not result in a conflict with $v1$, so there would be no need to look for alternative paths; the route length in this case would only be 6.
Thus, as long as a solution can be found without needing to change paths, \theAlgorithm will return an optimal solution, else optimality is not guaranteed.

\section{Conclusions} \label{sec:conclusions}

This paper presents the compositional algorithm \theAlgorithm to solve the \theModel. It is proven that the algorithm is sound complete if cycles in the paths are not allowed. \theAlgorithm  was compared to the performance of a monolithic model for the \theModel, which showed that as the problem sizes grow \theAlgorithm outperforms the monolithic model. \theAlgorithm's performance was also evaluated over a set of larger generated problem instances, which showed that it can solve problems counting up to 11 vehicles and 15 jobs in a reasonably short time. From the experimental data, it can be concluded that \theAlgorithm's solving time is subject to variability, depending on the problem's intrinsic complexity.

One advantage of \theAlgorithm, is that each sub-problem can be improved individually without affecting the complexity of the others. One possibility is to use different solvers for each sub-problem;
For the \emph{Routing Problem}, for example, MILP or specific purpose algorithms could replace the SMT formulation/solver, potentially resulting in increased performance. 

However, for problems whose feasibility is not hard to determine, i.e., problems where a reasonable amount of vehicles is available and the road segments have sufficient capacity, \theAlgorithm scales well, providing a schedule for rather large problems in a short time. This property is fulfilled in many industrial scenarios.

For the problems whose feasibility/infeasibility is not trivial to determine, it is an open research question on how to avoid a large number of iterations but at the same time do not increase the sub-problem complexity.

\bibliographystyle{IEEEtran}
\bibliography{bibliography}

\begin{IEEEbiography}[{\includegraphics[width=1in,height=1.25in,clip,keepaspectratio]{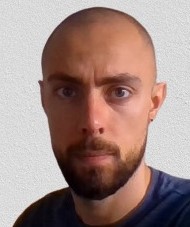}}]{Sabino Roselli} 
was born in Bari, Italy, in 1992. He received a M.Sc. degree in Industrial Engineering from Politecnico di Bari, Bari, Italy, in 2017. Since then, he  has  been  pursuing  a Ph.D.  degree  at the  Electrical  Engineering Department, Chalmers. His field of research is optimal scheduling of operations within the industrial context. 
\end{IEEEbiography}

 \begin{IEEEbiography}[{\includegraphics[height=1.25in,clip,keepaspectratio]{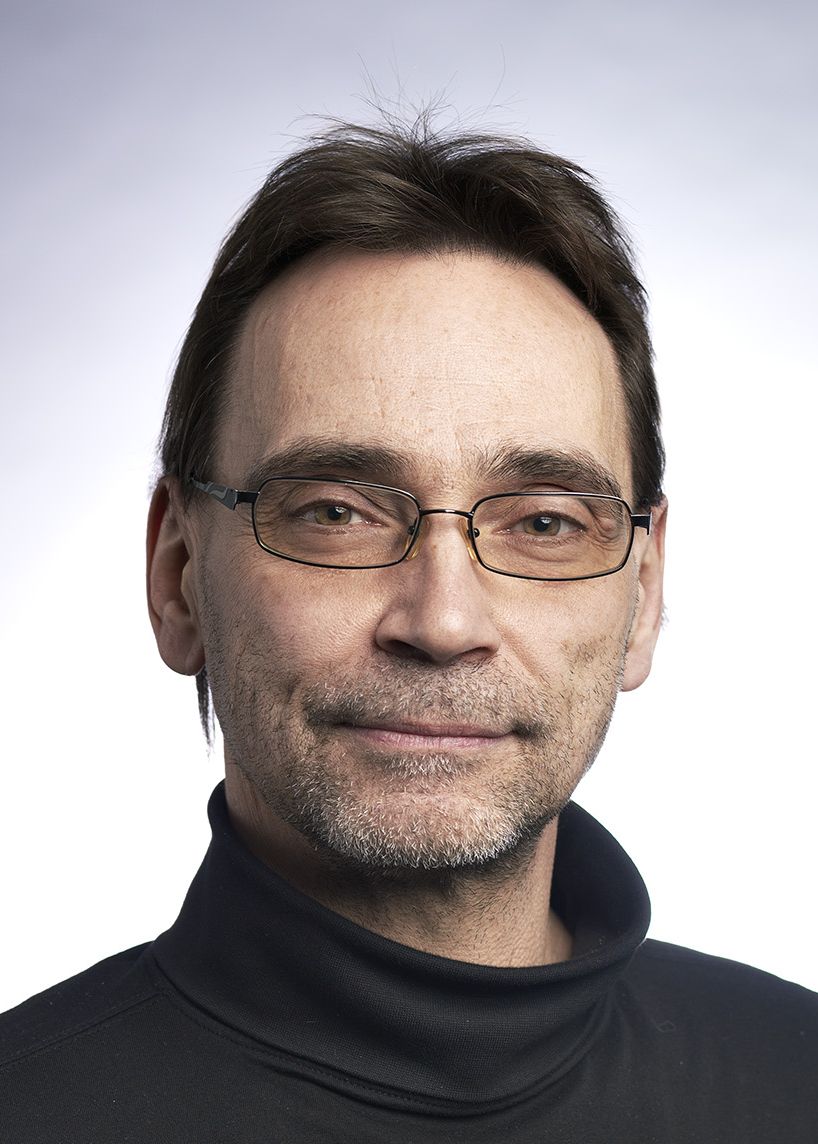}}]{Martin Fabian}
 is Full Professor in Automation and Head of the Automation Research group at the Department of Electrical Engineering, Chalmers University of Technology. His research interests include formal methods for automation systems in a broad sense, spanning the fields of Control Engineering and Computer Science. He has authored more than 200 publications, and is co-developer of the formal methods tool Supremica, which implements several state-of-the-art algorithms for supervisory control synthesis.
 \end{IEEEbiography}

\begin{IEEEbiography}[{\includegraphics[width=1in,height=1.2in,clip]{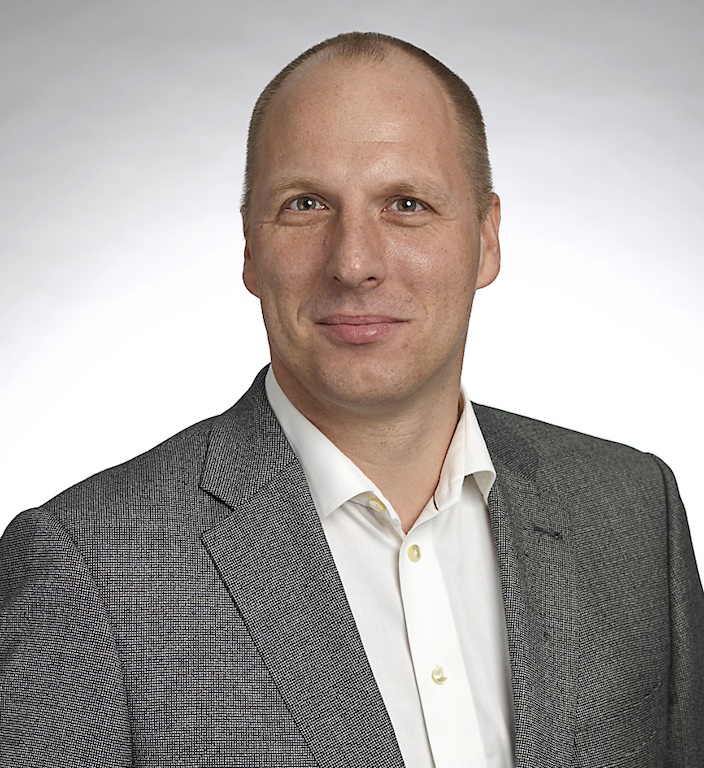}}]{Knut Åkesson}
is Professor in the Department of Electrical Engineering at Chalmers University of Technology, Gothenburg, Sweden. His main research is in using rigorous methods for analysis of cyber-physical systems. Åkesson holds a M.Sc. in Computer Science and Technology from Lund Institute of Technology, Sweden, and PhD in Control Engineering from Chalmers University of Technology, Gothenburg, Sweden.
\end{IEEEbiography}
\vfill

\begin{sidewaystable}[htbp]
  \centering
  \caption{Comparison of ComSat and the monolithic model (MonoMod) for the \theModel ~over a set of generated problem instances. Instances are sorted by the parameters N-V-J (nodes, vehicles, jobs), value of edge reduction, and time horizon. For each resulting class, five instances are evaluated and the number of feasible (Feas) and unfeasible (Unfeas) ones is reported, together with the average solving time (in seconds) for that specific class. When the number of instances does not add up to five it is because the running time exceeded the time limit of 1200 seconds. The symbol ``-'' means that no instance for that category was either feasible or unfeasible, depending on where the symbol appears.}

   \resizebox{\columnwidth}{!}{%

    \begin{tabular}{cc|cccc|cccc|cccc|cccc|cccc|cccc}
    \toprule
    \multicolumn{2}{c}{\multirow{3}[4]{*}{}} & \multicolumn{24}{c}{N-V-J} \\
\cmidrule{3-26}    \multicolumn{2}{c}{} & \multicolumn{12}{c}{15-3-5 }                              & \multicolumn{12}{c}{25-4-7 } \\
    \multicolumn{2}{c}{} & \multicolumn{12}{c}{Edge Reduction}                       & \multicolumn{12}{c}{Edge Reduction} \\
    \midrule
    T  &    & \multicolumn{4}{c}{0} & \multicolumn{4}{c}{25} & \multicolumn{4}{c|}{50} & \multicolumn{4}{c}{0} & \multicolumn{4}{c}{25} & \multicolumn{4}{c}{50} \\
      & & Av.(sec) & Feas & Av.(sec) & Unfeas & Av.(sec) & Feas & Av.(sec) & Unfeas & Av.(sec) & Feas & Av.(sec) & Unfeas & Av.(sec) & Feas & Av.(sec) & Unfeas & Av.(sec) & Feas & Av.(sec) & Unfeas & Av.(sec) & Feas & Av.(sec) & Unfeas \\
    \multirow{2}[0]{*}{20} & \cellcolor[rgb]{ .929,  .929,  .929}ComSat & \cellcolor[rgb]{ .929,  .929,  .929}1.74 & \cellcolor[rgb]{ .929,  .929,  .929}1 & \cellcolor[rgb]{ .929,  .929,  .929}120.43 & \cellcolor[rgb]{ .929,  .929,  .929}4 & \cellcolor[rgb]{ .929,  .929,  .929}0.63 & \cellcolor[rgb]{ .929,  .929,  .929}1 & \cellcolor[rgb]{ .929,  .929,  .929}106.75 & \cellcolor[rgb]{ .929,  .929,  .929}4 & \cellcolor[rgb]{ .929,  .929,  .929}-& \cellcolor[rgb]{ .929,  .929,  .929}0 & \cellcolor[rgb]{ .929,  .929,  .929}49.12 & \cellcolor[rgb]{ .929,  .929,  .929}5 & \cellcolor[rgb]{ .929,  .929,  .929}-& \cellcolor[rgb]{ .929,  .929,  .929}0 & \cellcolor[rgb]{ .929,  .929,  .929}539.04 & \cellcolor[rgb]{ .929,  .929,  .929}4 & \cellcolor[rgb]{ .929,  .929,  .929}-& \cellcolor[rgb]{ .929,  .929,  .929}0 & \cellcolor[rgb]{ .929,  .929,  .929}367.59 & \cellcolor[rgb]{ .929,  .929,  .929}4 & \cellcolor[rgb]{ .929,  .929,  .929}-& \cellcolor[rgb]{ .929,  .929,  .929}0 & \cellcolor[rgb]{ .929,  .929,  .929}284.82 & \cellcolor[rgb]{ .929,  .929,  .929}4 \\
       & MonoMod & 5.7 & 1 & 5.81 & 4 & 11.87 & 1 & 8.74 & 4 & - & 0 & 12.13 & 5 & - & 0 & 21.35 & 5 & - & 0 & 19.98 & 5 & - & 0 & 15.43 & 5 \\
    \multirow{2}[0]{*}{25} & \cellcolor[rgb]{ .929,  .929,  .929}ComSat & \cellcolor[rgb]{ .929,  .929,  .929}2.71 & \cellcolor[rgb]{ .929,  .929,  .929}2 & \cellcolor[rgb]{ .929,  .929,  .929}125.16 & \cellcolor[rgb]{ .929,  .929,  .929}3 & \cellcolor[rgb]{ .929,  .929,  .929}0.67 & \cellcolor[rgb]{ .929,  .929,  .929}2 & \cellcolor[rgb]{ .929,  .929,  .929}90.28 & \cellcolor[rgb]{ .929,  .929,  .929}3 & \cellcolor[rgb]{ .929,  .929,  .929}1.23 & \cellcolor[rgb]{ .929,  .929,  .929}1 & \cellcolor[rgb]{ .929,  .929,  .929}51.83 & \cellcolor[rgb]{ .929,  .929,  .929}4 & \cellcolor[rgb]{ .929,  .929,  .929}6.27 & \cellcolor[rgb]{ .929,  .929,  .929}1 & \cellcolor[rgb]{ .929,  .929,  .929}599.9 & \cellcolor[rgb]{ .929,  .929,  .929}3 & \cellcolor[rgb]{ .929,  .929,  .929}-& \cellcolor[rgb]{ .929,  .929,  .929}0 & \cellcolor[rgb]{ .929,  .929,  .929}315.15 & \cellcolor[rgb]{ .929,  .929,  .929}4 & \cellcolor[rgb]{ .929,  .929,  .929}-& \cellcolor[rgb]{ .929,  .929,  .929}0 & \cellcolor[rgb]{ .929,  .929,  .929}190.63 & \cellcolor[rgb]{ .929,  .929,  .929}4 \\
       & MonoMod & 10.51 & 3 & 235.28 & 2 & 16.08 & 3 & 13.52 & 1 & 26.63 & 1 & 217.15 & 3 & 100.16 & 1 & 63.24 & 4 & - & 0 & 44.34 & 5 & - & 0 & 39.35 & 5 \\
    \multirow{2}[0]{*}{30} & \cellcolor[rgb]{ .929,  .929,  .929}ComSat & \cellcolor[rgb]{ .929,  .929,  .929}6.72 & \cellcolor[rgb]{ .929,  .929,  .929}3 & \cellcolor[rgb]{ .929,  .929,  .929}0.36 & \cellcolor[rgb]{ .929,  .929,  .929}2 & \cellcolor[rgb]{ .929,  .929,  .929}3.73 & \cellcolor[rgb]{ .929,  .929,  .929}2 & \cellcolor[rgb]{ .929,  .929,  .929}179.19 & \cellcolor[rgb]{ .929,  .929,  .929}3 & \cellcolor[rgb]{ .929,  .929,  .929}2.78 & \cellcolor[rgb]{ .929,  .929,  .929}1 & \cellcolor[rgb]{ .929,  .929,  .929}89.03 & \cellcolor[rgb]{ .929,  .929,  .929}4 & \cellcolor[rgb]{ .929,  .929,  .929}15.3 & \cellcolor[rgb]{ .929,  .929,  .929}2 & \cellcolor[rgb]{ .929,  .929,  .929}804.06 & \cellcolor[rgb]{ .929,  .929,  .929}2 & \cellcolor[rgb]{ .929,  .929,  .929}4.19 & \cellcolor[rgb]{ .929,  .929,  .929}1 & \cellcolor[rgb]{ .929,  .929,  .929}410.97 & \cellcolor[rgb]{ .929,  .929,  .929}3 & \cellcolor[rgb]{ .929,  .929,  .929}98.25 & \cellcolor[rgb]{ .929,  .929,  .929}1 & \cellcolor[rgb]{ .929,  .929,  .929}200.49 & \cellcolor[rgb]{ .929,  .929,  .929}3 \\
       & MonoMod & 25.71 & 3 & 4.45 & 1 & 80.33 & 3 & 9.14 & 1 & 112.24 & 1 & 57.64 & 2 & 292.1 & 2 & 145.42 & 2 & 242.48 & 2 & 70.34 & 2 & 159.43 & 1 & 242.91 & 3 \\
    \multirow{2}[0]{*}{40} & \cellcolor[rgb]{ .929,  .929,  .929}ComSat & \cellcolor[rgb]{ .929,  .929,  .929}1.75 & \cellcolor[rgb]{ .929,  .929,  .929}3 & \cellcolor[rgb]{ .929,  .929,  .929}0.37 & \cellcolor[rgb]{ .929,  .929,  .929}2 & \cellcolor[rgb]{ .929,  .929,  .929}6.16 & \cellcolor[rgb]{ .929,  .929,  .929}3 & \cellcolor[rgb]{ .929,  .929,  .929}0.37 & \cellcolor[rgb]{ .929,  .929,  .929}2 & \cellcolor[rgb]{ .929,  .929,  .929}3.83 & \cellcolor[rgb]{ .929,  .929,  .929}3 & \cellcolor[rgb]{ .929,  .929,  .929}0.37 & \cellcolor[rgb]{ .929,  .929,  .929}2 & \cellcolor[rgb]{ .929,  .929,  .929}5.23 & \cellcolor[rgb]{ .929,  .929,  .929}4 & \cellcolor[rgb]{ .929,  .929,  .929}671.25 & \cellcolor[rgb]{ .929,  .929,  .929}1 & \cellcolor[rgb]{ .929,  .929,  .929}192.15 & \cellcolor[rgb]{ .929,  .929,  .929}4 & \cellcolor[rgb]{ .929,  .929,  .929}462.68 & \cellcolor[rgb]{ .929,  .929,  .929}1 & \cellcolor[rgb]{ .929,  .929,  .929}4.98 & \cellcolor[rgb]{ .929,  .929,  .929}3 & \cellcolor[rgb]{ .929,  .929,  .929}190.19 & \cellcolor[rgb]{ .929,  .929,  .929}1 \\
       & MonoMod & 342.44 & 3 & 9.41 & 1 & 162.11 & 2 & 34.59 & 1 & 210.6 & 2 & 19.71 & 1 & 403.96 & 2 & - & 0 & 534.69 & 2 & - & 0 & 414.29 & 2 & - & 0 \\
    \multirow{2}[0]{*}{50} & \cellcolor[rgb]{ .929,  .929,  .929}ComSat & \cellcolor[rgb]{ .929,  .929,  .929}0.74 & \cellcolor[rgb]{ .929,  .929,  .929}3 & \cellcolor[rgb]{ .929,  .929,  .929}0.39 & \cellcolor[rgb]{ .929,  .929,  .929}2 & \cellcolor[rgb]{ .929,  .929,  .929}4.99 & \cellcolor[rgb]{ .929,  .929,  .929}3 & \cellcolor[rgb]{ .929,  .929,  .929}0.37 & \cellcolor[rgb]{ .929,  .929,  .929}2 & \cellcolor[rgb]{ .929,  .929,  .929}2.04 & \cellcolor[rgb]{ .929,  .929,  .929}3 & \cellcolor[rgb]{ .929,  .929,  .929}0.36 & \cellcolor[rgb]{ .929,  .929,  .929}2 & \cellcolor[rgb]{ .929,  .929,  .929}5.08 & \cellcolor[rgb]{ .929,  .929,  .929}4 & \cellcolor[rgb]{ .929,  .929,  .929}964.66 & \cellcolor[rgb]{ .929,  .929,  .929}1 & \cellcolor[rgb]{ .929,  .929,  .929}7.66 & \cellcolor[rgb]{ .929,  .929,  .929}4 & \cellcolor[rgb]{ .929,  .929,  .929}466.4 & \cellcolor[rgb]{ .929,  .929,  .929}1 & \cellcolor[rgb]{ .929,  .929,  .929}14.54 & \cellcolor[rgb]{ .929,  .929,  .929}3 & \cellcolor[rgb]{ .929,  .929,  .929}345.57 & \cellcolor[rgb]{ .929,  .929,  .929}2 \\
       & MonoMod & 237.64 & 3 & 16.14 & 1 & 412.25 & 3 & 15.89 & 1 & 704.12 & 3 & 20.49 & 1 & 497.04 & 1 & - & 0 & 628.88 & 1 & - & 0 & - & 0 & - & 0 \\
    \multirow{2}[1]{*}{60} & \cellcolor[rgb]{ .929,  .929,  .929}ComSat & \cellcolor[rgb]{ .929,  .929,  .929}1.58 & \cellcolor[rgb]{ .929,  .929,  .929}3 & \cellcolor[rgb]{ .929,  .929,  .929}0.36 & \cellcolor[rgb]{ .929,  .929,  .929}2 & \cellcolor[rgb]{ .929,  .929,  .929}6.46 & \cellcolor[rgb]{ .929,  .929,  .929}3 & \cellcolor[rgb]{ .929,  .929,  .929}0.37 & \cellcolor[rgb]{ .929,  .929,  .929}2 & \cellcolor[rgb]{ .929,  .929,  .929}8.01 & \cellcolor[rgb]{ .929,  .929,  .929}3 & \cellcolor[rgb]{ .929,  .929,  .929}0.36 & \cellcolor[rgb]{ .929,  .929,  .929}2 & \cellcolor[rgb]{ .929,  .929,  .929}4.29 & \cellcolor[rgb]{ .929,  .929,  .929}5 & \cellcolor[rgb]{ .929,  .929,  .929}-& \cellcolor[rgb]{ .929,  .929,  .929}0 & \cellcolor[rgb]{ .929,  .929,  .929}15.02 & \cellcolor[rgb]{ .929,  .929,  .929}5 & \cellcolor[rgb]{ .929,  .929,  .929}-& \cellcolor[rgb]{ .929,  .929,  .929}0 & \cellcolor[rgb]{ .929,  .929,  .929}5.2 & \cellcolor[rgb]{ .929,  .929,  .929}5 & \cellcolor[rgb]{ .929,  .929,  .929}-& \cellcolor[rgb]{ .929,  .929,  .929}0 \\
       & MonoMod & 462.53 & 2 & - & 0 & 553.51 & 2 & - & 0 & 742.8 & 2 & - & 0 & - & 0 & - & 0 & - & 0 & - & 0 & - & 0 & - & 0 \\
    \bottomrule
    \end{tabular}%

    }
    
  \label{tab:comparison_table}%
\end{sidewaystable}%

\begin{sidewaystable}[htbp]
  \centering
  \caption{Evaluation of ComSat for the \theModel ~over a set of generated problem instances. Instances are sorted by the parameters N-V-J (nodes, vehicles, jobs), and value of edge reduction in the columns, and time horizon in the rows.
  For each resulting class, five instances are evaluated and the number of feasible (Feas) and unfeasible (Unfeas) ones is reported, together with the average solving time (in seconds) for that specific class.
  When the number of instances does not add up to five it is because the running time exceeded the time limit of 1200 seconds. The symbol ``-'' means that no instance for that category was either feasible or unfeasible, depending on where the symbol appears.}
  
    \resizebox{\columnwidth}{!}{%

    \begin{tabular}{c|cccc|cccc|cccc|cccc|cccc|cccc}
    \toprule
    \multicolumn{1}{c}{\multirow{3}[4]{*}{}} & \multicolumn{24}{c}{N-V-J} \\
\cmidrule{2-25}    \multicolumn{1}{c}{} & \multicolumn{12}{c}{35-6-8 }                              & \multicolumn{12}{c}{35-07-10} \\
    \multicolumn{1}{c}{} & \multicolumn{12}{c}{Edge Reduction}                       & \multicolumn{12}{c}{Edge Reduction} \\
    \midrule
    T  & \multicolumn{4}{c}{0} & \multicolumn{4}{c}{25} & \multicolumn{4}{c|}{50} & \multicolumn{4}{c}{0} & \multicolumn{4}{c}{25} & \multicolumn{4}{c}{50} \\
       & Av.(sec) & Feas & Av.(sec) & Unfeas & Av.(sec) & Feas & Av.(sec) & Unfeas & Av.(sec) & Feas & Av.(sec) & Unfeas & Av.(sec) & Feas & Av.(sec) & Unfeas & Av.(sec) & Feas & Av.(sec) & Unfeas & Av.(sec) & Feas & Av.(sec) & Unfeas \\
    \rowcolor[rgb]{ .929,  .929,  .929} 40 & 98.32 & 3 & 292.11 & 2 & 42.93 & 2 & 383.28 & 3 & 10.86 & 2 & 351.63 & 3 & 5.49 & 1 & 1.33 & 2 & -  & 0 & 1.33 & 2 & -  & 0 & 403.6 & 3 \\
    70 & 79.55 & 3 & 269.74 & 2 & 19.89 & 2 & 0.86 & 3 & 42.98 & 2 & 0.86 & 3 & 238.65 & 3 & 1.32 & 2 & 127.31 & 1 & 1.33 & 2 & 4.59 & 1 & 1.31 & 2 \\
    \rowcolor[rgb]{ .929,  .929,  .929} 100 & 15.35 & 3 & 89.12 & 2 & 79.9 & 2 & 281.04 & 3 & 73.86 & 2 & 281.66 & 3 & 452.1 & 3 & 1.31 & 2 & 22.01 & 1 & 1.33 & 2 & 7.48 & 1 & 1.31 & 2 \\
    150 & 8.94 & 3 & 100.53 & 2 & 47.33 & 3 & 0.87 & 2 & 56.84 & 2 & 0.86 & 3 & 28.44 & 3 & 1.3 & 2 & 16.79 & 1 & 1.39 & 2 & 32.41 & 1 & 1.31 & 2 \\
    \rowcolor[rgb]{ .929,  .929,  .929} 200 & 6.48 & 4 & 0.86 & 1 & 23.19 & 2 & 351.15 & 3 & 102.8 & 2 & 350.46 & 3 & 202.68 & 3 & 1.32 & 2 & 66.64 & 1 & 1.34 & 2 & 21.18 & 1 & 1.29 & 2 \\
    300 & 8.99 & 4 & 0.88 & 1 & 34.1 & 3 & 0.86 & 2 & 17.21 & 2 & 0.86 & 3 & 74.12 & 3 & 1.34 & 2 & 6.36 & 1 & 1.33 & 2 & 79.27 & 1 & 1.35 & 2 \\
    \midrule
    \multicolumn{1}{c}{\multirow{3}[4]{*}{}} & \multicolumn{24}{c}{N-V-J} \\
\cmidrule{2-25}    \multicolumn{1}{c}{} & \multicolumn{12}{c}{35-9-12 }                             & \multicolumn{12}{c}{35-11-15} \\
    \multicolumn{1}{c}{} & \multicolumn{12}{c}{Edge Reduction}                       & \multicolumn{12}{c}{Edge Reduction} \\
    \midrule
    T  & \multicolumn{4}{c}{0} & \multicolumn{4}{c}{25} & \multicolumn{4}{c|}{50} & \multicolumn{4}{c}{0} & \multicolumn{4}{c}{25} & \multicolumn{4}{c}{50} \\
       & Av.(sec) & Feas & Av.(sec) & Unfeas & Av.(sec) & Feas & Av.(sec) & Unfeas & Av.(sec) & Feas & Av.(sec) & Unfeas & Av.(sec) & Feas & Av.(sec) & Unfeas & Av.(sec) & Feas & Av.(sec) & Unfeas & Av.(sec) & Feas & Av.(sec) & Unfeas \\
    \rowcolor[rgb]{ .929,  .929,  .929} 40 & 486.31 & 2 & 1.6 & 3 & 311.95 & 1 & 1.88 & 2 & -  & 0 & 1.9 & 3 & -  & 0 & 3.17 & 1 & -  & 0 & 3.19 & 4 & -  & 0 & 3.24 & 4 \\
    70 & 221.11 & 3 & 1.89 & 1 & 107.44 & 1 & 321.05 & 2 & 143.61 & 1 & 1.94 & 3 & -  & 0 & 3.26 & 1 & -  & 0 & 3.33 & 4 & -  & 0 & 3.32 & 4 \\
    \rowcolor[rgb]{ .929,  .929,  .929} 100 & 77.16 & 2 & 1.88 & 2 & 28.02 & 1 & 1.9 & 2 & 111.09 & 1 & 1.92 & 3 & 437.57 & 2 & -  & 0 & 341.98 & 1 & 3.17 & 3 & -  & 0 & 3.2 & 3 \\
    150 & 224.31 & 3 & 1.92 & 2 & 255.06 & 2 & 1.95 & 2 & 311.23 & 2 & 1.99 & 3 & 337.26 & 2 & -  & 0 & -  & 0 & 3.22 & 3 & -  & 0 & 3.26 & 3 \\
    \rowcolor[rgb]{ .929,  .929,  .929} 200 & 489.57 & 3 & 1.89 & 2 & 27.59 & 1 & 1.88 & 2 & 178.64 & 1 & 1.94 & 3 & 18.38 & 1 & 3.19 & 2 & 788.91 & 1 & 3.27 & 4 & 661.28 & 1 & 3.22 & 4 \\
    300 & 238.88 & 3 & 1.92 & 2 & 391.58 & 1 & 1.92 & 2 & 349.04 & 1 & 1.94 & 3 & 646.29 & 2 & -  & 0 & 774.71 & 1 & 3.26 & 3 & 26.26 & 1 & 3.31 & 3 \\
    \bottomrule
    \end{tabular}%
    
    }
    
  \label{tab:performance_table}%
\end{sidewaystable}%

\end{document}